\documentclass[11pt]{article} 
\usepackage{fullpage}
\usepackage{todonotes}
\usepackage[T1]{fontenc}
\usepackage[utf8]{inputenc}
\usepackage[english]{babel}
\newcommand\ppsrule[1]{\textit{\sc#1}}
\newcommand\ruleres{\ppsrule{Resolution}}
\newcommand\rulesim{\ppsrule{Simplify}}
\newcommand\ruleintro{\ppsrule{Introduce}}
\newcommand\ruleweak{\ppsrule{Weak}}
\newcommand\rulecl{\ppsrule{Clause}}
\newcommand\trres{\ppsrule{$\odot$-Res}}
\newcommand\pps[1]{\textbf{#1}}
\newcommand\rescp{\pps{Res(CP)}}
\newcommand\reslp{\pps{Res(LP)}}
\newcommand\mpnsr{\pps{MP-NSR}}
\newcommand\mppcr{\pps{MP-PCR}}
\newcommand\mpns{\pps{MP-NS}}
\newcommand\mppc{\pps{MP-PC}}
\newcommand\thrres{\pps{Thr-Res}}
\newcommand{\LnP}{\pps{L\&P}}
\newcommand{\CP}{\pps{CP}}
\newcommand{\res}{\pps{Res}}
\usepackage{lmodern}
\usepackage{amsmath,amssymb,amsthm,mathtools}
\usepackage{graphicx,subcaption}
\usepackage{microtype}
\usepackage{xcolor}
\usepackage[colorlinks=true,linkcolor=blue,citecolor=blue,urlcolor=blue]{hyperref}
\let\uprightamp\&
\DeclareRobustCommand{\&}{\textup{\uprightamp}}
\newcommand{\Q}{\mathbb{Q}}
\newcommand{\Z}{\mathbb{Z}}
\newcommand{\poly}{\operatorname{poly}}
\theoremstyle{plain}
\newtheorem{theorem}{Theorem}
\newtheorem{lemma}[theorem]{Lemma}
\newtheorem{proposition}[theorem]{Proposition}
\newtheorem{corollary}[theorem]{Corollary}
\theoremstyle{definition}
\newtheorem{definition}[theorem]{Definition}
\theoremstyle{remark}
\newtheorem{example}[theorem]{Example}
\newtheorem{remark}[theorem]{Remark}
\usepackage{chngcntr}
\usepackage{tikz}
\tikzset{
  sys/.style={draw, rounded corners, fill=gray!10, inner sep=4pt, align=center, font=\footnotesize},
  sim/.style={->, >=stealth, semithick},
  equ/.style={<->, >=stealth, semithick},
  lab/.style={font=\scriptsize, fill=white, inner sep=1.5pt},
}
\usepackage{xspace}
\counterwithin{table}{section}
\counterwithin{figure}{section}
\counterwithin{theorem}{section}
\newcommand{\coef}{\operatorname{coef}}

\begin{document}

\title{Further simulations between tropical and linear programming\thanks{Department of Computer Science,
School of Computer Science and Artificial Intelligence,
Ariel University, Israel.
Email: \url{edwardh@ariel.ac.il}, \url{dmitrybsluch@gmail.com}. This research was supported by The Israel Science Foundation (grant No. 1645/26).}} %
\author{Edward A. Hirsch
\and
Dmitrii Sluch}
\date{November 18, 2026}
\maketitle

\begin{abstract}
Propositional proof systems that apply various kinds of cuts to linear inequalities
(such as \pps{Cutting Planes} \cite{CCT87} and \pps{Lift-and-Project} \cite{BCC93,Pud99,Dash05,HK06}) or operate with disjunctions
of linear inequalities (such as {\rescp} \cite{Kra98} and {\reslp} \cite{HK06}) have been in the focus of
research for the past 30 years. Recently, tropical proof systems were
introduced \cite{AGH25}; they resemble the latter systems but differ from them in important respects.
The relations between the traditional systems and the new tropical systems have remained largely unknown, and exponential lower bounds for tropical systems were limited to non-CNF examples.

  In this paper we prove several new polynomial simulations between these systems.
To this end, we introduce {\thrres} (Threshold Resolution), a new natural daglike proof system operating on threshold conditions over the Boolean literals: lines asserting that at least $s$ literals of a given multiset are true. Its main difference from {\CP}
is that the rounding rule (without it {\CP} is incomplete) is replaced by a weaker rule
that allows one to perform ``resolution'' steps with linear combinations beyond Boolean clauses;
also, in its addition rules, {\thrres} resembles \emph{input resolution} \cite{Chang70}: a derived line is combined with an input clause or an axiom,
whereas {\CP} can add up any two derived lines.
By varying the maximum value of $s$, one obtains systems ranging in strength from \pps{Resolution} to {\mpnsr}, the static tropical (min-plus) proof system introduced in \cite{AGH25}.

As a consequence, we prove that {\mpnsr} is polynomially simulated by
the \pps{Cutting Planes} and \pps{Lift-and-Project} proof systems (even by their unary versions), which automatically yields
exponential lower bounds on the size of {\mpnsr} %
proofs for formulas in CNF. 
We complement this result with simulations in the converse direction: binary daglike {\mppcr} (a dynamic version of {\mpnsr}, also introduced in \cite{AGH25}) polynomially simulates \pps{Lift-and-Project};
it also polynomially simulates \pps{Cutting Planes} when division axioms are added.
\end{abstract}

\pagebreak\tableofcontents\pagebreak %

\section{Introduction}
\label{sec:intro}

\paragraph{Motivation and background.} Propositional proof complexity asks about the existence of short proofs of propositional tautologies (or short refutations of unsatisfiable propositional formulas): given an unsatisfiable CNF formula, a \emph{refutation} in a specific proof system is a certificate of unsatisfiability that can be checked in polynomial time (a proof system is associated with such a complete and sound proof verification algorithm). The central question is to show that for each system there is a family of formulas admitting only superpolynomially long refutations. Comparing two systems is done by \emph{simulation}: a system $\Pi_2$ \emph{polynomially simulates} $\Pi_1$ if every $\Pi_1$-refutation can be converted into a $\Pi_2$-refutation of polynomially related size. Simulations organize proof systems into a lattice, and locating a new system in this lattice is the standard way to understand its strength. Standard definitions in the field have been established in the foundational paper by Cook and Reckhow~\cite{CR79}. The first
and best-known propositional proof system is propositional \emph{Resolution} ({\res}) \cite{Rob65,DP60}:
starting from the input clauses, one is allowed to derive new clauses using the resolution rule
$\displaystyle\frac{C\lor x,\ D\lor \overline{x}^{\phantom{I}}}{C\lor D_{\phantom{I}}}$ until a contradiction (the empty clause)
is derived; exponential lower bounds were first proved for regular Resolution \cite{Tse68} and then for general Resolution \cite{Hak85,Urq87}; they have been known for decades. It is still the basis of the most competitive
CDCL SAT solvers \cite{MSS99,MMZZM01,BN21}.

A number of proof systems considered in the literature have been motivated by research in combinatorial optimization: for example, proof systems \cite{Pud99,GV01,G02,GHP02,BS14,DM12,G01parity} motivated by the Lov\'asz-Schrijver, Sherali-Adams, and Lasserre cuts \cite{LS91,SA90,Las01}. One of the most classical systems is the \emph{Cutting Planes} proof system ({\CP}) \cite{CCT87} based on the Gomory-Chv\'atal cuts \cite{Gom58,Chv73}. Exponential lower bounds have been known for it since \cite{Kra97,Pudlak97},
while this question remains open for its extension {\rescp} \cite{Kra98}
that deals with disjunctions of linear inequalities. A treelike version of the latter system,
called \emph{Stabbing Planes} \cite{Stabbing,Stabbing2}, has received much attention recently.
Subexponential lower bounds were proved for it
by Gl\"aser and Pfetsch 
\cite{GlasserPfetsch24}, and
quasipolynomial upper bounds on the length of refutations of Tseitin formulas
were proved in \cite{Stabbing} and then transferred to {\CP}
by Dadush and Tiwari \cite{DT20}.
Another system interesting for us is Lift-and-Project ({\LnP}) \cite{BCC93};
an exponential lower bound for it was proved by Dash in \cite{Das01,Dash05}.
We define these systems below when we come to the details.

Alekseev, Grigoriev and Hirsch~\cite{AGH25} introduced proof systems based on tropical (min-plus) arithmetic, where multiplication is the ordinary addition and addition is the minimum: the static system {\mpns}, based on the tropical (min-plus) Nullstellensatz of Grigoriev and Podolskii~\cite{GrPo18}, exhibits a single contradictory combination of the premises, while its dynamic counterpart {\mppc} derives new inequalities between tropical polynomials step by step
(see \cite{JM,BE,ABG} for other tropical Nullstellens\"atze). Their Boolean versions {\mpnsr} and {\mppcr} refute CNF formulas encoded by tropical inequalities. 

At first glance, {\mpnsr} may look like a weak system,
because of its \emph{static} nature (that is, a proof in it is a collection of 
tropical terms, and its correctness is verified by computing a single polynomial being
a sum-product combination of the input polynomials with the proof terms --- very similar
in nature to proofs based on Hilbert's Nullstellensatz \cite{BeameEtAl96}).
However, it can polynomially simulate daglike Resolution proofs
and is exponentially stronger than Resolution, because
it has polynomial proofs for the propositional pigeon-hole principle \cite{AGH25}.

Tropical polynomials over Boolean variables are minima of affine functions, so tropical reasoning is also a form of reasoning about systems of linear inequalities. 
While \cite{AGH25} proved a polynomial equivalence between
{\reslp} (which is a simplified version of {\rescp})
and {\mppcr} extended by an additional rule,
little has been known about further connections.
In particular, relations between {\mpnsr} and {\CP} (another system
that both simulates {\res} and has short proofs of the propositional
pigeon-hole principle) were unknown.

\medskip
In this paper we resolve this question by showing that {\mpnsr}
can be polynomially simulated in {\CP} (as well as in {\LnP})
and show additional polynomial simulations between systems based 
on systems of linear inequalities and those based on tropical arithmetic
(thus mixing linear inequalities with disjunctions).
Our results also imply the first exponential lower bounds for {\mpnsr}
for formulas in CNF.
The basis of our new results is a new proof system {\thrres}.

\paragraph{Basic proof systems.}
Before we come to the new proof system, we briefly introduce
known proof systems that are important for us. In what follows,
$x_1, \ldots, x_N$ denote variables, $x=(x_1,\ldots,x_N)$.
See Section~\ref{sec:prelim}
for formal definitions.

The \emph{Cutting Planes} ({\CP}) \cite{CCT87}
and \emph{Lift-and-Project} ({\LnP}) \cite{BCC93,Pud99,Dash05,HK06}
proof systems operate with linear inequalities
and derive the contradictory inequality $0\ge1$
as a proof of infeasibility of the original system of inequalities.
To convert formulas in CNF into systems of inequalities, a Boolean clause
is converted into the corresponding inequality,
for instance, $x_1 \vee x_2 \vee \overline{x_3}$ becomes $x_1 + x_2 + (1 - x_3) \geq 1$; 
the axioms $x_i \geq 0$ and $x_i \leq 1$ (the \emph{box axioms}) are added for each variable $x_i$.

Both systems allow one to derive nonnegative linear combinations of
previously derived inequalities. The difference is cuts
(here $c \in \Z^N$ and $d,d',d'' \in \Z$):
\begin{itemize}
\item
{\CP} uses the rounding rule that forces the variables to be integers:
from $d'\, c^T x \geq d$, where $d' > 0$, 
one can derive $c^T x \geq \lceil d/d' \rceil$,
\item
{\LnP} uses lift-and-project cuts: from $c^T x + d' \geq d$ and $c^T x + d'' x_k \geq d$,
one can derive $c^T x + d' x_k \geq d$.
\end{itemize}

\emph{Tropical proof systems} \cite{AGH25} are based on tropical arithmetic:
the tropical addition $\oplus$ is the minimum,
the tropical multiplication $\odot$ is the ``normal'' addition.
The constants are taken from $\mathbb{Q}\cup\{+\infty\}$,
while the degrees remain nonnegative integers.
This way one can form tropical polynomials
in a way similar to the usual polynomials.
A tropical term (a monomial with a coefficient)
corresponds to an affine form, and
the conventional meaning of a tropical inequality
between two tropical polynomials
is a conjunction of disjunctions of inequalities:
note that
$f_1\oplus\ldots\oplus f_{\ell}\le g$ is equivalent
to $\bigvee_{i=1}^{\ell} (f_i\le g)$,
and $f\le g_1\oplus\ldots\oplus g_r$ is equivalent
to $\bigwedge_{i=1}^r (f\le g_i)$.

Static tropical proof systems are based
on a tropical Nullstellensatz \cite{GrPo18}.
An {\mpns} refutation of a system of tropical inequalities
takes these inequalities, possibly augmenting this list with
repetitions and \emph{trivial axioms}
($0 \leq 0$ or $p \leq \infty$ for any tropical polynomial $p$),
and multiplies each inequality $h_j\le h_j'$ in the resulting list $h_1 \leq h_1', \ldots, h_K \leq h_K'$
by a tropical term $t_j$ so that the inequality
$
\bigoplus_{j=1}^{K} t_j \odot h_{j}
\;\leq\;
\bigoplus_{j=1}^{K} t_j \odot h_{j}'
$
becomes \emph{contradictory}, that is,
the right-hand side has a finite constant term and,
for every monomial $m$ occurring in the left-hand side as some term $c\odot m$,
its $c$ is strictly greater than the coefficient of $m$ in the right-hand side.
The size of the refutation is $K$ plus the total bitsize of the coefficients of the terms $t_j$ plus the total bitsize of all exponents.

A dynamic version of this system,
{\mppc}, derives new tropical inequalities line by line.
A derivation from the system of inequalities that we refute
is a sequence of tropical inequalities $p_1 \leq q_1, \ldots, p_K \leq q_K$, 
each of which is either an original inequality, a trivial axiom, or is obtained from 
earlier inequalities by one of the rules:
\begin{itemize}
\item \ppsrule{Minimum:} from $p \leq q$ and $p' \leq q'$ derive $p \oplus p' \leq q \oplus q'$;
\item \ppsrule{Tropical multiplication:} from $p \leq q$ derive $p \odot t \leq q \odot t$, where $t$ is a tropical term;
\item \ppsrule{Transitivity:} from $p \leq h$ and $h \leq r$ derive $p \leq r$.
\end{itemize}
A derivation is a \emph{refutation} if its last line is contradictory 
(in fact, one can always finish the proof by a contradiction $0\ge c$ for a positive constant $c$).

Boolean versions of tropical systems ({\mpnsr} and {\mppcr}, respectively)
introduce two \emph{dual} variables $x_i$, $\overline{x_i}$
for each variable $x_i$ of the input Boolean formula
and add the axioms
\begin{align*}
&0 \leq x_i, \qquad 0 \leq \overline{x_i} && \text{(the weakening axioms)},\\
&1 \leq x_i \odot \overline{x_i}, \qquad x_i \odot \overline{x_i} \leq 1 && \text{(the pair axioms)},\\
&x_i \oplus \overline{x_i} \leq 0 && \text{(the resolution axiom)}.
\end{align*}
A clause $C$ is encoded by the tropical inequality $1 \leq \bigodot_{l \in C} l$, the tropical form of its linear inequality above with $\overline{x_i}$ in place of $1 - x_i$; for instance, $x_1 \vee x_2 \vee \neg x_3$ becomes $1 \leq x_1 \odot x_2 \odot \overline{x_3}$.

\emph{The unary and the binary settings.}
All these proof systems can be considered in the \emph{binary} setting
where the coefficients and exponents are written in binary
and in the \emph{unary} setting
where they are written in unary
(for tropical systems, it applies
to numerators and denominators of the rationals).
It affects the size of the proof,
in particular, the values of the coefficients and exponents
in a polynomial-size proof in the unary setting
are polynomially bounded.
It is an open question to prove a superpolynomial
separation between these settings for {\CP};
it is only known that larger-than-exponential
coefficients are not needed in the binary setting:
such proofs can be always rewritten
using coefficients of polynomial bit-size \cite{CCT87,BC96},
the only known separation requires
space-bounded setting \cite{dRMNPRV20}.
In contrast, for {\mpnsr} and treelike {\mppcr} refutations of CNFs, the unary and binary settings are polynomially equivalent (Corollaries~\ref{cor:unary-coefficients} and~\ref{cor:triangle-standard}).
Unless stated otherwise, tropical proof systems use binary encoding of coefficients and individual exponents, whereas {\thrres} uses unary encoding.

\paragraph{A new proof system {\thrres} and our results.}
This paper introduces a daglike proof system on linear inequalities and locates the tropical systems with its help. {\thrres} (Threshold Resolution) operates on \emph{threshold conditions} over the Boolean literals --- lines asserting that at least $s$ literals of a given multiset are true --- with arbitrary integer thresholds. The \emph{degree of falsity} of a {\thrres} proof is the maximum threshold occurring in it. Restricting this parameter gives the \emph{{\thrres} degree-of-falsity hierarchy}. For every fixed constant $k\geq1$, {\thrres} restricted to proofs of degree of falsity at most $k$ is polynomially equivalent to daglike Resolution. The central result of the paper, the \emph{Main Lemma}, identifies unrestricted unary {\thrres} with the binary static tropical system: binary {\mpnsr} and {\thrres} are polynomially equivalent. The hierarchy thus interpolates between daglike Resolution and binary {\mpnsr}. For {\mpnsr} refutations of CNFs, both coefficients and exponents can be made unary at polynomial cost (Corollary~\ref{cor:unary-coefficients}).
Combining the Main Lemma with the simulation of {\thrres} by unary Cutting Planes and Lift \& Project, we obtain that unary Cutting Planes and Lift \& Project polynomially simulate {\mpnsr}. This yields
the first exponential lower bounds for {\mpnsr} for formulas in CNF
(before that, they were known only for systems of equations not related to CNFs: tropical binomials such as $x_1^{\odot 2^n}=-1$, whose exponent is exponentially large and written in binary~\cite[Theorem~7.16]{AGH25}).
In the converse direction, binary daglike {\mppcr} simulates daglike Lift \& Project and, with division axioms added, daglike Cutting Planes. Figure~\ref{fig:lattice} summarizes these simulations.
We also give a quantitative simulation of bounded-degree {\thrres} by Resolution (Theorem~\ref{thm:constant-k-collapse}) and bounds within the hierarchy for the pigeonhole principle and Tseitin formulas (Propositions~\ref{prop:php-hierarchy} and~\ref{prop:tseitin-degree}). The equivalence between the static, treelike tropical and daglike threshold systems extends uniformly to arbitrary unary classes of tropical axioms (Appendix~\ref{sec:appendix-generalized}).

\begin{figure}[htp]
\centering
\begin{tikzpicture}[x=1cm, y=1cm]
\node[sys] (res)    at (1.0, 0)   {daglike Resolution}; %
\node[sys] (thr1)   at (6.0, 0)   {{\thrres}\\ degree of falsity $\leq 1$};
\node[sys] (thrk)   at (6.0, 1.5) {{\thrres}\\ degree of falsity $\leq k$, $k\geq2$};
\node[sys] (thrinf) at (6.0, 3.0) {{\thrres}};
\node[sys] (mpnsr)  at (11.0, 3.0) {binary\\ {\mpnsr}}; %
\node[sys] (tmppcr) at (14.0, 3.0) {binary treelike\\ {\mppcr}}; %
\node[sys] (udlnp)  at (2.6, 5.0) {unary daglike {\LnP}};
\node[sys] (udcp)   at (8.4, 5.0) {unary daglike {\CP}}; %
\node[sys] (dlnp)   at (2.6, 6.6) {daglike {\LnP}}; %
\node[sys] (dcp)    at (8.4, 6.6) {daglike {\CP}}; %
\node[sys] (mppcrres) at (2.6, 8.3) {binary daglike {\mppcr}}; %
\node[sys] (mppcrdiv) at (8.4, 8.3) {binary daglike {\mppcr}\\ with the division axioms};
\draw[equ] (res) -- (thr1) node[lab, midway, above=2pt] {Theorem~\ref{thm:1-thrres-resolution}};
\draw[sim] (thr1) -- (thrk);
\draw[sim] (thrk) -- (thrinf);
\draw[equ] (thrinf) -- (mpnsr) node[lab, midway, above] {Main Lemma \ref{lm:main}};
\draw[equ] (mpnsr) -- (tmppcr) node[lab, midway, above=12pt] {\cite{AGH25}};
\draw[sim] (thrinf) to[bend right=20] node[lab, below] {Theorem~\ref{thm:daglike-to-treelike}} (tmppcr);
\draw[sim] (thrinf) -- (udlnp) node[lab, midway, above, sloped] {Lemma~\ref{lm:lp-cp-simulate-thrres}};
\draw[sim] (thrinf) -- (udcp) node[lab, midway, above, sloped] {Lemma~\ref{lm:lp-cp-simulate-thrres}};
\draw[sim] (udlnp) -- (dlnp);
\draw[sim] (udcp) -- (dcp);
\draw[sim] (dlnp) -- (mppcrres) node[lab, midway, right] {Theorem~\ref{thm:mppcr-simulates-lp}};
\draw[sim] (dcp) -- (mppcrdiv) node[lab, midway, right] {Theorem~\ref{thm:mppcr-simulates-cp}};
\draw[sim] (mppcrres) -- (mppcrdiv);
\end{tikzpicture}
\caption{The proof systems studied in this paper. An arrow from $\Pi_1$ to $\Pi_2$ means that $\Pi_2$ polynomially simulates $\Pi_1$ ($\Pi_1 \leq \Pi_2$).
The direct proof of the simulation from {\thrres} to treelike {\mppcr} (Theorem~\ref{thm:daglike-to-treelike}) is redundant for establishing the equivalence of binary {\mpnsr}, binary treelike {\mppcr} and unary {\thrres} (Corollary~\ref{cor:triangle-standard}), but is included for its simplicity. Both this proof and the proof of the Main Lemma extend to arbitrary unary classes of tropical axioms (Appendix~\ref{sec:appendix-generalized}). All {\thrres} systems in the diagram are unary. Both {\mppcr} systems of the upper part include the standard and the resolution axioms; the right one additionally includes the division axioms. Unlabeled arrows hold by definition; every other arrow is labeled by the statement that proves it. For every constant $k$, {\thrres} restricted to degree of falsity at most $k$ collapses to Resolution (Theorem~\ref{thm:constant-k-collapse}, not shown).}  %
\label{fig:lattice}
\end{figure}

\paragraph{Our methods.}
The proof of the Main Lemma goes through a graph-theoretic reading of binary {\mpnsr} certificates. A certificate determines a finite weighted graph on monomial positions: each position selects a premise to apply, its outgoing edges encode the resulting obligations, and the global consistency condition is that every directed cycle has strictly positive weight. Unlike a proof, this graph may contain cycles and records only differences between thresholds, rather than thresholds themselves. Layering the graph by accumulated weight resolves both difficulties: positivity of the cycles allows a polynomial-size acyclic unfolding whose local transitions form a backward {\thrres} proof. Conversely, we compress a backward proof by retaining an occurrence with the maximum threshold at each position, then reconstruct a static certificate using vertex potentials. The construction remains polynomial even for binary certificates: the number of graph vertices is polynomially bounded independently of the exponent magnitudes, paths from the origin bound the degrees of their monomial labels, and the bounded edge weights yield unary thresholds. This argument extends uniformly to arbitrary unary classes of tropical axioms (Appendix~\ref{sec:appendix-generalized}).

The polynomial equivalence of binary {\mpnsr} and binary treelike {\mppcr} was established in~\cite{AGH25}; we reuse its flattening direction below. A short alternative proof of the direction from {\thrres} to binary {\mpnsr} first translates a daglike threshold proof into a treelike {\mppcr} proof (Theorem~\ref{thm:daglike-to-treelike}), then applies the flattening theorem of~\cite[Theorem~4.1]{AGH25}. The translation from {\thrres} to treelike {\mppcr} accumulates all processed threshold lines in a single tropical inequality and applies each input premise in its rule context. It also extends to arbitrary unary classes of tropical axioms, and the same premise-in-context pattern reappears in the comparison with Lift \& Project.

The simulations of {\thrres} by {\CP} and {\LnP} translate threshold conditions into linear inequalities and simulate each rule locally. In the reverse direction, the Lift \& Project step is represented by a derivable tropical inequality applied in a common context, while binary scaling is implemented efficiently by repeated squaring in daglike {\mppcr}. To simulate {\CP}, we additionally use division axioms to represent its rounding rule.

\paragraph{Organization of the paper.}
The paper is organized as follows. Section~\ref{sec:prelim} defines the tropical systems (previously known) and Threshold Resolution (our new system). Section~\ref{sec:main-results} states the Main Lemma and proves the polynomial equivalence of binary {\mpnsr} and {\thrres} by the graph-theoretic argument, then gives the daglike-to-treelike simulation and records the resulting cycle of simulations. Section~\ref{sec:lp-connection} establishes the simulations between tropical systems, {\CP} and {\LnP}, and derives lower bounds for the tropical systems. Section~\ref{sec:hierarchy} studies the degree-of-falsity hierarchy, including the quantitative simulation by daglike Resolution and the bounds for the pigeonhole principle and Tseitin formulas within the degree-of-falsity hierarchy. Section~\ref{sec:open} collects open questions. Appendix~\ref{sec:appendix-generalized} extends the equivalence to arbitrary unary classes of tropical axioms.

\section{Preliminaries}
\label{sec:prelim}

\subsection{Notation and derivation types}
\label{sec:notation}
Throughout the paper, $N$ denotes the number of variables in the input formula and $M$ the number of input constraints (clauses, inequalities) --- the input clause encodings as well as the axioms standard for the system; $[n]$ denotes $\{1, \ldots, n\}$. Variables are indexed as $x_1, \ldots, x_N$, with $i \in [N]$. For a rational $q = \frac{n}{d}$ in lowest terms with $d \geq 1$, the \emph{bitsize} of $q$ is $\lceil\log_2(|n| + 1)\rceil + \lceil\log_2(d + 1)\rceil + 1$ (so that the bitsize of $0$ is well defined and positive). We use $S$ for the size of a proof, $\deg$ for degrees of tropical monomials, and $W$ for bounds on the absolute values of coefficients and constants.

Traditionally, proof systems are classified by their structural properties:
\begin{itemize}
\item \textbf{daglike} systems derive new lines from already derived lines until a contradiction is reached; a proof forms a directed acyclic graph, and a derived line may be used many times;
\item in \textbf{treelike} systems a proof forms a tree instead of an arbitrary dag, i.e., each derived line may be used only once, so reuse requires re-derivation; 
\item \textbf{static} systems exhibit a single certificate (e.g., a collection of multipliers); a static refutation is verified by checking one global combination of the premises.
\end{itemize}
For many systems operating with numbers, both \emph{unary} and \emph{binary} versions are traditionally considered: in unary variants, integer coefficients, thresholds, and degrees are encoded in unary and their unary encoding lengths are included in the proof size; for ``binary'' versions, the size of the proof assumes binary encoding of these numbers. Consequently, every polynomial-size unary proof contains only polynomially bounded numerical values. Unless stated otherwise, tropical proof systems use binary encoding of coefficients and individual exponents, whereas {\thrres} uses unary encoding.
\subsection{Tropical polynomials}
\label{sec:tropical-arithmetic}

Tropical (min-plus) arithmetic replaces addition by the minimum and multiplication by the ordinary addition: we write $a \oplus b = \min(a, b)$ and $a \odot b = a + b$, over $\Q_\infty := \Q \cup \{+\infty\}$, with $+\infty$ the tropical zero. Following~\cite{AGH25}, the semantic domain is $\Q_\infty^n$: variables may take the value $+\infty$. A \emph{tropical monomial} in the variables $y_1, \ldots, y_n$ is a formal product $y^I := \bigodot_{j=1}^{n} y_j^{I_j}$ with $I \in \Z_+^{n}$, which evaluates at $y \in \Q_\infty^n$ to $\langle I, y\rangle := \sum_j I_j y_j$, using $0\cdot(+\infty)=0$ and $a\cdot(+\infty)=+\infty$ for positive integers $a$; we identify monomials with their exponent vectors $I$, write $\deg y^I = \sum_j I_j$, and denote by $0$ the constant monomial (the zero vector). A \emph{tropical term} is a monomial with a coefficient, $c \odot y^I$, evaluating to the affine function $c + \langle I, y\rangle$. A \emph{tropical polynomial} is a formal tropical sum
\[
f \;=\; \bigoplus_{I} c_I \odot y^{I}, \qquad c_I \in \Q_\infty ,
\]
with finitely many finite coefficients; it evaluates at a point $y \in \Q_\infty^n$ to $\min_{I}\big(c_I + \langle I, y\rangle\big)$, the minimum of its terms. We write $\coef_m(f) := c_m$ for the coefficient of the monomial $m$ in $f$; a monomial \emph{occurs} in $f$ if $\coef_m(f) < +\infty$. We freely use \emph{non-reduced presentations} of tropical polynomials: several terms with the same monomial may be displayed separately, although only the smallest of their coefficients affects the represented polynomial --- for instance, $1 \oplus (-1)$ and $-1$ are two presentations of the same tropical polynomial.

A \emph{tropical inequality} is a pair $(f, g)$ of tropical polynomials, written $f \leq g$; it holds at a point $y \in \Q_\infty^n$ if $f(y) \leq g(y)$ under tropical evaluation. A system of tropical inequalities $\{f_i \leq g_i : i \in [M]\}$, the \emph{premises}, is \emph{unsatisfiable} if no point of $\Q_\infty^n$ satisfies all of them. Since $f \leq g_1 \oplus \cdots \oplus g_r$ is equivalent to the collection of the inequalities $f \leq g_j$, $j \in [r]$, we may assume where convenient that an inequality has a single term on the right-hand side.

\subsection{Contradictory inequalities}
\label{sec:contradictory}

The tropical systems below refute a system of premises by combining them into a single inequality that fails everywhere in a syntactically evident way.

\begin{definition}[Contradictory inequality]
\label{def:contradictory}
For tropical polynomials $p$ and $q$, write $p \succ q$ if
\[
\coef_m(p) \;>\; \coef_m(q) \qquad \text{for every monomial } m \text{ occurring in } p,
\]
and the constant coefficient $\coef_0(q)$ of $q$ is finite. A tropical inequality $p \leq q$ is \emph{contradictory} if $p \succ q$.
\end{definition}

Note that $\coef_m(p) > \coef_m(q)$ forces $\coef_m(q)$ to be finite, so every monomial of $p$ occurs in $q$ with a strictly smaller coefficient. A contradictory inequality fails at every point of the semantic domain. Let $y \in \Q_\infty^n$. The finite constant coefficient of $q$ guarantees $q(y) \leq \coef_0(q) < +\infty$. Consider a term of $p$ indexed by a monomial $m$. If $\langle m, y\rangle = +\infty$, then this term is strictly larger than $q(y)$. Otherwise,
\[
q(y) \;\leq\; \coef_m(q) + \langle m, y\rangle \;<\; \coef_m(p) + \langle m, y\rangle .
\]
Thus every term of $p$ is strictly larger than $q(y)$, and hence $p(y) > q(y)$.

\begin{example}
The inequality $1 \odot y_1 \;\leq\; y_1 \oplus 0$ (that is, $y_1 + 1 \leq \min(y_1, 0)$) is contradictory: the only monomial of the left-hand side is $y_1$, with coefficient $1 > 0 = \coef_{y_1}(q)$, and $\coef_0(q) = 0$ is finite. The inequality $1 \odot y_1 \leq y_1$ satisfies the coefficient condition but is \emph{not} contradictory: the right-hand side has no finite constant term, and indeed the inequality holds at the point $y_1 = +\infty$; the finite-constant condition of Definition~\ref{def:contradictory} is exactly what excludes such parasite solutions with coordinates equal to $+\infty$.
\end{example}

\subsection{MP-NS and MP-PC}
\label{sec:mp-systems}

We now define the two basic tropical proof systems of Alekseev, Grigoriev and Hirsch~\cite{AGH25}: the static {\mpns} and the dynamic {\mppc}. Both refute a system of premises
\[
\mathcal{S} \;=\; \{f_i \leq g_i : i \in [M]\},
\]
and both may additionally use the \emph{trivial axioms}
\[
0 \leq 0 \qquad\text{and}\qquad p \leq \infty \quad \text{(for any tropical polynomial } p\text{)},
\]
which hold everywhere and are available besides the finite input list. These trivial axiom schemas are included in both the general systems and their Boolean versions below.

\begin{definition}[MP-NS]
\label{def:mp-ns}
An {\mpns} refutation of $\mathcal{S}$ is a finite list of inequalities $h_1 \leq h_1', \ldots, h_K \leq h_K'$, each of which is a premise of $\mathcal{S}$ or a trivial axiom, together with tropical terms $t_1, \ldots, t_K$, such that the inequality
\begin{equation}
\label{eq:mp-ineq}
\bigoplus_{j=1}^{K} t_j \odot h_{j}
\;\leq\;
\bigoplus_{j=1}^{K} t_j \odot h_{j}'
\end{equation}
is contradictory (Definition~\ref{def:contradictory}). The size of the refutation is $K$ plus the total bitsize of the coefficients of the terms $t_j$ plus the total bitsize of all exponents. Unless stated otherwise, we use this binary-exponent size measure for {\mpns} and {\mpnsr}; for {\mpnsr} refutations of CNFs, both exponents and coefficients can be made unary at polynomial cost even from a binary-encoded certificate (Corollary~\ref{cor:unary-coefficients}), but not for arbitrary systems of premises (Remark~\ref{rem:unary-beyond-cnf}).
\end{definition}

Each side of~\eqref{eq:mp-ineq} is a finite tropical sum of tropical multiples of the premises, so a refutation is a single algebraic combination of the premises, exhibited at once --- {\mpns} is a static system. It is sound: at any common point of the premises each $t_j \odot h_{j}$ evaluates to at most $t_j \odot h_{j}'$, hence the left minimum is at most the right one, contradicting that~\eqref{eq:mp-ineq} fails everywhere. It is also complete for unsatisfiable systems of tropical inequalities: this is the Min-Plus Nullstellensatz, due to Grigoriev and Podolskii~\cite[Theorem~3.8]{GrPo18}; see \cite[Theorems~5 and~6]{AGH25} for the form over $\Q_\infty$ used here.

\begin{example}
\label{ex:mp-ns}
The premises $\{y_1 \leq 0,\ 1 \leq y_1\}$ are unsatisfiable. Taking each premise once with the term $t = 0$ gives the combination $y_1 \oplus 1 \leq 0 \oplus y_1$, which is \emph{not} contradictory: the monomial $y_1$ has the coefficient $0$ on both sides, and the definition requires strict inequality. Multiplying the first premise by $t = \tfrac{1}{2}$ instead gives $\tfrac12 \odot y_1 \oplus 1 \;\leq\; \tfrac12 \oplus y_1$: now the coefficients of $y_1$ on the two sides are $\tfrac12 > 0$, the constant coefficients are $1 > \tfrac12$, and the constant $\tfrac12$ on the right is finite --- a contradictory combination, hence an {\mpns} refutation. The example also shows why non-integer coefficients are natural in tropical refutations.
\end{example}

\begin{definition}[MP-PC]
\label{def:mp-pc}
An {\mppc} derivation from $\mathcal{S}$ is a sequence of tropical inequalities $p_1 \leq q_1, \ldots, p_K \leq q_K$, each of which is a premise of $\mathcal{S}$, a trivial axiom, or is obtained from earlier lines by one of the rules:
\begin{itemize}
\item \textbf{Minimum:} from $p \leq q$ and $p' \leq q'$ derive $p \oplus p' \leq q \oplus q'$;
\item \textbf{Tropical multiplication:} from $p \leq q$ derive $p \odot t \leq q \odot t$, where $t$ is a tropical term;
\item \textbf{Transitivity:} from $p \leq h$ and $h \leq r$ derive $p \leq r$.
\end{itemize}
A derivation is a \emph{refutation} if its last line is contradictory (Definition~\ref{def:contradictory}). The derivation is \emph{treelike} if every derived line is used as a premise of a rule at most once. The size of a derivation is the number of lines plus the total bitsize of all coefficients and the total bitsize of all exponents in it; in the unary version, coefficients and exponents are written in unary.
\end{definition}

\begin{remark}[Substitutions]
\label{rem:substitutions}
Combining transitivity with the other rules, one can substitute derived inequalities into one another on the left or on the right (\cite[Note~2.12]{AGH25}): from $p \oplus h \leq q$ and $r \leq h$ derive $p \oplus r \leq q$, and from $p \leq q \oplus h$ and $h \leq r$ derive $p \leq q \oplus r$. We use substitutions without further notice.
\end{remark}

\subsection{Boolean encoding and the resolution axioms}
\label{sec:boolean-encoding}

To refute CNF formulas, each Boolean variable $x_i$ comes together with a complementary variable $\overline{x_i}$; the two are separate tropical variables, so tropical polynomials over them live in $2N$ variables. A \emph{literal} $l$ is one of $x_i$, $\overline{x_i}$; the complement $\overline{l}$ of a literal is the other one. The intended semantics is $x_i \in \{0, 1\}$ and $\overline{x_i} = 1 - x_i$, with $1$ read as \emph{true}; it is enforced by the following axioms.

\begin{definition}[{\mpnsr} and {\mppcr}]
\label{def:mpnsr}
{\mpnsr} (resp.\ {\mppcr}) is {\mpns} (resp.\ {\mppc}) over the dual-encoded variables with the following axioms added to the premises, for every $i \in [N]$:
\begin{align*}
&0 \leq x_i, \qquad 0 \leq \overline{x_i} && \text{(the weakening axioms)},\\
&1 \leq x_i \odot \overline{x_i}, \qquad x_i \odot \overline{x_i} \leq 1 && \text{(the pair axioms)},\\
&x_i \oplus \overline{x_i} \leq 0 && \text{(the resolution axiom)}.
\end{align*}
We call the weakening and pair axioms jointly the \emph{standard axioms}, and $\min(x_i, \overline{x_i}) \leq 0$ the \emph{resolution axiom}.
\end{definition}

At a point of $\Q_\infty^{2N}$ satisfying all five axioms, $\min(x_i, \overline{x_i}) = 0$ (the weakening axioms give $\geq$, the resolution axiom gives $\leq$) and $x_i + \overline{x_i} = 1$ (the pair axioms) --- in particular both values are finite --- so one of $x_i, \overline{x_i}$ equals $0$ and the other equals $1$: the common roots are exactly the Boolean points. This is the axiom set of~\cite{AGH25} with $0 \leq x_i \oplus \overline{x_i}$ split into the two weakening axioms: in {\mppcr}, the split axioms yield the combined one by minimum, and each split axiom follows by dropping a right-hand summand. In {\mpnsr}, replace each use of the combined axiom by both split axioms with the same multiplier, leaving the combination unchanged; conversely, replacing a split axiom by the combined one can only lower right-hand coefficients, preserving contradictoriness.

\begin{remark}[Eliminating trivial axioms from static refutations]
\label{rem:trivial-static}
{\mpnsr} with and without the trivial axiom schemas are polynomially equivalent as refutation systems; indeed, the trivial entries can be deleted from any {\mpns} refutation without increasing its size. Let $P\leq Q$ be its contradictory combination. Entries using $p\leq\infty$ contribute nothing to $Q$. An entry using $0\leq0$ with multiplier $\lambda\odot m$ cannot attain $\coef_m(Q)$, since otherwise $\coef_m(P)\leq\lambda=\coef_m(Q)$, contradicting $P\succ Q$. Deleting all trivial entries therefore leaves $Q$ unchanged and only raises left-hand coefficients or removes left-hand monomials, preserving contradictoriness. This is equivalence for refutations, not implicational equivalence: without the trivial schemas, a nonempty combination of the Boolean premises always has a finite right-hand term and hence cannot equal $p\leq\infty$.
\end{remark}

\begin{remark}[The resolution axiom, variable resolution, and {\trres}]
\label{rem:odot-res}
The resolution axiom
\[
x_i \oplus \overline{x_i} \;\leq\; 0
\]
is part of the Boolean encoding used by {\mpnsr} and {\mppcr}. Applied inside a threshold context (in the sense made precise in Lemma~\ref{lm:rules-are-applications}), it yields the variable-resolution rule
\[
\frac{f + x_i \geq s \qquad f + \overline{x_i} \geq s}{f \geq s}\,,
\]
the {\ruleres} rule of the threshold system defined in Section~\ref{sec:thrres-def}. This rule should be distinguished from the substantially more general inference rule {\trres} of~\cite{AGH25},
\[
\frac{t \oplus f \leq 0 \qquad t' \oplus f \leq 0}{(t \odot t') \oplus f \leq 0}\,, \qquad \text{where $t$ and $t'$ are tropical terms,}
\]
a counterpart of the addition rule of $\mathrm{{\reslp}}$~\cite{HK06} and {\rescp}~\cite{Kra98}: it combines two arbitrary affine functions $t$ and $t'$ inside a common tropical context $f$, rather than merely resolving a variable against its complement. The rule {\trres} is \emph{not} part of {\mppcr}, {\mpnsr}, or the threshold systems as defined in this paper; it is not obtained here from the resolution axiom, and it appears substantially stronger, since it operates on arbitrary linear combinations --- \cite{AGH25} add it to their systems explicitly when needed and show that its elimination may be expensive.
\end{remark}

\begin{definition}[Clause encoding]
\label{def:clause-encoding}
A clause $C$ of a CNF formula $F$ is encoded by the tropical inequality
\begin{equation}
1 \;\leq\; \bigodot_{l \in C} l .
\tag{I}
\label{eq:encoding-i}
\end{equation}
For example, the clause $x_1 \vee x_2 \vee \neg x_3$ becomes $1 \leq x_1 \odot x_2 \odot \overline{x_3}$, that is, $1 \leq x_1 + x_2 + \overline{x_3}$. An {\mpnsr} (or {\mppcr}) refutation of $F$ is a refutation of the system consisting of the encodings of the clauses of $F$ and the axioms of Definition~\ref{def:mpnsr}.
\end{definition}

The encoding~\eqref{eq:encoding-i} states that at least one literal of the clause is true; it is the standard encoding of clauses in {\CP}-like systems. It is the only clause encoding used in this paper.

\begin{remark}[The dual encoding]
\label{rem:encoding-d}
A clause can alternatively be encoded by $\bigoplus_{l \in C} \overline{l} \leq 0$ (``some complemented literal is false''). The two encodings are polynomially equivalent: there are polynomial-size unary treelike {\mppcr} derivations of each from the other (\cite{AGH25} Lemma~3.5), so any use of the dual encoding may be treated as a derived rule. The dual encoding transfers in the same way to the threshold system of Section~\ref{sec:thrres-def}: read as an inference rule in the manner of Lemma~\ref{lm:rules-are-applications} --- from the premises $f + \overline{l} \geq s$, one for every $l \in C$, derive $f \geq s$ --- it is interderivable with the {\rulecl} rule of {\thrres} (defined in Section~\ref{sec:thrres-def} below) and may be used there as a derived rule as well.

To see this, let $C=\{l_1,\ldots,l_r\}$ be nonempty. From $f+\overline{l_1}\geq s$, {\rulecl} followed by {\rulesim} gives $f+\sum_{j=2}^r l_j\geq s$. For $j=2,\ldots,r$, weaken the premise $f+\overline{l_j}\geq s$ by $\sum_{q=j+1}^r l_q$ and resolve against the current line, successively obtaining $f+\sum_{q=j+1}^r l_q\geq s$ and finally $f\geq s$.
Conversely, from $f\geq s$, {\ruleintro} on $l_j,\overline{l_j}$ followed by {\ruleweak} gives $f+\sum_{l\in C}l+\overline{l_j}\geq s+1$ for every $j$; applying the dual rule yields $f+\sum_{l\in C}l\geq s+1$, the {\rulecl} conclusion. Both constructions have polynomial size; an empty input clause is already a contradiction.
\end{remark}

\subsection{A new system: Threshold Resolution}
\label{sec:thrres-def}

The daglike threshold system that in its unrestricted form will be shown equivalent to {\mpnsr} operates on \emph{threshold conditions}: linear inequalities
\[
f \geq s, \qquad f = \sum_{l} a_l\, l, \quad a \in \Z_+^{2N},\ s \in \Z,
\]
where the sum ranges over the $2N$ literals; a literal and its complement may occur in $f$ simultaneously, and literals occur with multiplicities. This is the threshold form $c^T x \geq d$ of LP-based proof systems, restricted to nonnegative coefficients over the dual-encoded variables. We write $|f| = \sum_l a_l$ for the number of literals of $f$ counted with multiplicity; for example, $|x_i + \overline{x_i}| = 2$. Over the Boolean points, $f \geq s$ states that at least $s$ of the literals of $f$ (with multiplicity) are true; in particular $f \geq 1$ is the disjunction of the literals of $f$, and $f \geq |f|$ is their conjunction.

\begin{definition}[Threshold Resolution]
\label{def:thrres}
A proof in the system \emph{{\thrres}} (Threshold Resolution) is a finite directed acyclic graph in which:
\begin{enumerate}
\item Every vertex is labeled by a threshold condition $f \geq s$ with an arbitrary integer threshold $s$; the labels are the \emph{lines} of the proof.
\item There is a single source, labeled $0 \geq 0$.
\item Every non-source vertex, together with all its incoming edges --- one from each premise of the rule --- represents an application of one of the rules below, deriving its line from the lines of its predecessors.
\end{enumerate}
In the rules below, $f$ is an arbitrary nonnegative integer linear combination of literals and $s$ is an arbitrary integer.
\begin{itemize}
\item {\rulecl} (clause introduction): from $f \geq s$ derive $f + \sum_{l \in C} l \geq s + 1$, for a clause $C$ of the input CNF. This introduces the encoding~\eqref{eq:encoding-i} of $C$ into the context.
\item {\ruleres}: from $f + x_i \geq s$ and $f + \overline{x_i} \geq s$ derive $f \geq s$.
\item {\ruleweak}: from $f \geq s$ derive $f + l \geq s$, for any literal $l$.
\item {\ruleintro}: from $f \geq s$ derive $f + l + \overline{l} \geq s + 1$.
\item {\rulesim}: from $f + l + \overline{l} \geq s$ derive $f \geq s - 1$.
\end{itemize}
These rules essentially allow adding Boolean axioms and input clauses to any threshold condition (context); this reading is made precise in Lemma~\ref{lm:rules-are-applications}.

A \emph{refutation} is a proof containing the line $0 \geq 1$. The size of a proof is the number of lines plus the total length of all coefficients and thresholds encoded in unary; the system is unary in the sense of Section~\ref{sec:notation}, so a polynomial-size proof contains only polynomially bounded coefficients and thresholds.
\end{definition}

\begin{remark}[Duplicate-free proofs]
\label{rem:no-duplicates}
A {\thrres} proof is \emph{duplicate-free} if no two vertices carry the same threshold condition $f\geq s$. Every proof can be made duplicate-free: once a line is derived, there is no need to derive it again. Indeed, fix a topological order, redirect the outgoing edges of all vertices carrying a given label to the earliest vertex with that label, and delete the other copies together with any vertices left hanging. The size does not increase, and the graph stays acyclic, since the source of every redirected edge only moves backwards in the topological order. From now on we consider only duplicate-free proofs.
\end{remark}

\begin{example}[Resolution in context]
\label{ex:resolution-rule}
Consider an instance of the {\ruleres} rule with the threshold fixed to $s = 1$ (the rule itself allows an arbitrary threshold): it derives $f \geq 1$ from $x_i + f \geq 1$ and $\overline{x_i} + f \geq 1$. Over the Boolean points this is a case analysis: if $x_i = 0$, the first premise yields $f \geq 1$; if $x_i = 1$, then $\overline{x_i} = 0$ and the second premise yields $f \geq 1$. Note that simply adding the corresponding linear inequalities is not enough: substituting $\overline{x_i} = 1 - x_i$, the sum of the premises is
\[
(x_i + f) + (1 - x_i + f) \;\geq\; 2, \qquad\text{that is,}\qquad 2f \;\geq\; 1 ,
\]
and Cutting Planes must further divide by $2$ and round, using the integrality of the variables, to reach $f \geq 1$; so $2f \geq 1$ is only an intermediate inequality, not the resolution conclusion. The bare tropical operations --- minimum and addition of a term --- likewise do not perform this case split. In the tropical setting the split is provided by the resolution axiom $x_i \oplus \overline{x_i} \leq 0$, whose application in the context $f - 1$ gives this instance directly, and in the context $f - s$ the general rule (Remark~\ref{rem:odot-res} and Lemma~\ref{lm:rules-are-applications}).
\end{example}

\begin{example}
\label{ex:toy-thrres}
Let $F = x_1 \wedge \overline{x_1}$, encoded by $1 \leq x_1$ and $1 \leq \overline{x_1}$. The four lines
\[
\ell_1 = (0 \geq 0), \qquad \ell_2 = (x_1 \geq 1), \qquad \ell_3 = (\overline{x_1} \geq 1), \qquad \ell_4 = (0 \geq 1)
\]
form a {\thrres} refutation: $\ell_2$ and $\ell_3$ are {\rulecl} steps from the source, and $\ell_4$ is a {\ruleres} step from $\ell_2$ and $\ell_3$. The proof DAG is pictured below, with each application labeled by its rule and context.
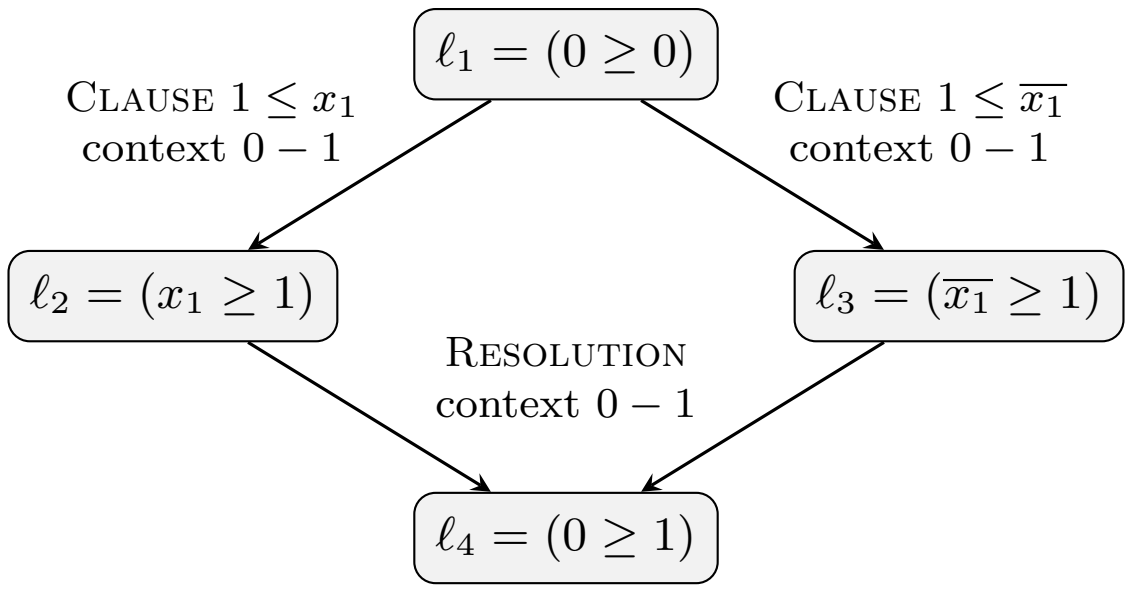
\begin{center}
\begin{tikzpicture}[x=1cm, y=1cm]
\node[sys] (l1) at (0, 0)  {$\ell_1 = (0 \geq 0)$};
\node[sys] (l2) at (-2.6, -1.6) {$\ell_2 = (x_1 \geq 1)$};
\node[sys] (l3) at (2.6, -1.6) {$\ell_3 = (\overline{x_1} \geq 1)$};
\node[sys] (l4) at (0, -3.2) {$\ell_4 = (0 \geq 1)$};
\draw[sim] (l1) -- (l2) node[lab, midway, above left, align=center] {{\rulecl} $1 \leq x_1$\\ context $0 - 1$};
\draw[sim] (l1) -- (l3) node[lab, midway, above right, align=center] {{\rulecl} $1 \leq \overline{x_1}$\\ context $0 - 1$};
\draw[sim] (l2) -- (l4);
\draw[sim] (l3) -- (l4);
\node[lab, align=center] at (0, -2.15) {{\ruleres}\\ context $0 - 1$};
\end{tikzpicture}
\end{center}
\end{example}

Some intuition for the rules. {\ruleweak} adds an arbitrary literal to a line: a form of weakening. {\ruleintro} and {\rulesim} are bookkeeping for the dual encoding: in a system encoding $\overline{x_i}$ as $1 - x_i$, the lines $f + x_i + \overline{x_i} \geq s + 1$ and $f \geq s$ would be one and the same inequality, and these two rules let us pass between the different ways of writing it. When translating a threshold condition into a linear inequality over the original Boolean variables, we cancel complementary pairs. The rules {\rulesim} and {\ruleintro} allow us to pass between these representations.

\begin{definition}[Degree of falsity]
\label{def:degree-falsity}
The \emph{degree of falsity} of a {\thrres} proof $\pi$ is
\[
\operatorname{df}(\pi):=\max\{s : (f\geq s)\text{ is a line of }\pi\}.
\]
For an integer $k\geq1$, restricting to proofs of degree of falsity at most $k$ means requiring $s\leq k$ at every vertex, including all intermediate lines. In particular, {\ruleintro} and {\rulecl} may be applied only if the resulting threshold is at most $k$. The refutation in Example~\ref{ex:toy-thrres} has degree of falsity $1$.
\end{definition}

The parameter measures how far the proof strays from clause-like reasoning: every nontrivial line of a proof of degree of falsity at most $1$ can be read as a disjunction, and this restriction of {\thrres} turns out to be polynomially equivalent to daglike Resolution (Theorem~\ref{thm:1-thrres-resolution}).

\begin{remark}[Degree of falsity for Cutting Planes]
\label{rem:hkkn}
The name of the parameter comes from the \emph{degree of falsity} of Goerdt~\cite{Goerdt91}, studied further by Hirsch and Nikolenko~\cite{DBLP:conf/sat/HirschN05} for Cutting Planes and by Hirsch, Kojevnikov, Kulikov and Nikolenko~\cite{DBLP:journals/jsat/HirschKKN09} for semialgebraic proofs. There, an inequality over the Boolean variables is normalized as $\sum_i a_i x_i + \sum_i b_i (1 - x_i) \geq c$ with $a_i, b_i \geq 0$, and its degree of falsity is its constant term $c$ (\cite[Lemma~1]{DBLP:conf/sat/HirschN05}). In the normalized literal form, at most one of $a_i,b_i$ is nonzero for each $i$. For a {\thrres} line $f\geq s$, write $f=\sum_i(a_i x_i+b_i\overline{x_i})$ and let $q=\sum_i\min(a_i,b_i)$ be the number of complementary pairs counted with multiplicity. Substituting $\overline{x_i}=1-x_i$ and cancelling these pairs gives normalized degree of falsity $s-q\leq s$, with equality when $q=0$. Thus a bound on the displayed thresholds bounds the normalized degrees of the translated lines. The extra intermediate lines needed for the {\CP} simulation are bounded separately in Theorem~\ref{thm:constant-k-collapse}.
\end{remark}

\begin{theorem}[Soundness]
\label{thm:thrres-sound}
Every line of a {\thrres} proof is satisfied by every Boolean point satisfying the input CNF. In particular, if a refutation exists, the CNF is unsatisfiable.
\end{theorem}
\begin{proof}
Induction in a topological order of the proof DAG; let $\alpha$ be a Boolean point satisfying the CNF. The source $0 \geq 0$ holds. For {\rulecl}, $\alpha$ satisfies some literal of $C$, hence $\sum_{l \in C} l \geq 1$ at $\alpha$, and adding this to the premise $f \geq s$ gives the conclusion. For {\ruleres}, one of $x_i, \overline{x_i}$ evaluates to $0$ at $\alpha$, and the corresponding premise gives $f \geq s$. For {\ruleweak}, the added literal is nonnegative. For {\ruleintro} and {\rulesim}, a complementary pair contributes exactly $1$ at $\alpha$.
\end{proof}

\section{{\thrres} and binary {\mpnsr} are polynomially equivalent}
\label{sec:main-results}

We first prove the Main Lemma by a graph-theoretic argument. We then give the simulation by treelike {\mppcr} and combine it with the flattening theorem of~\cite{AGH25}, obtaining the equivalence triangle and an alternative proof of the direction from {\thrres} to {\mpnsr}.

\subsection{The Main Lemma: statement and proof strategy}
\label{sec:main-lemma}
\label{sec:main-lemma-proof}

\begin{lemma}[Main Lemma]
\label{lm:main}
Binary {\mpnsr} and unary {\thrres} are polynomially equivalent. Quantitatively: a binary {\mpnsr} refutation of size $S$ of the encoding of a CNF $F$ yields a {\thrres} refutation of $F$ of size $\poly(S, N)$, and a {\thrres} refutation of size $S$ yields an {\mpnsr} refutation of size $\poly(S)$.
\end{lemma}

\begin{figure}[ht]
\centering
\begin{subfigure}{\textwidth}
\centering
\begin{tikzpicture}[x=1cm, y=1cm]
\node[sys] (r) at (0, 0) {$\vec 0$\\ \scriptsize {\ruleres} on $x_1$};
\node[sys] (a) at (-2.4, -1.6) {$e(x_1)$\\ \scriptsize {\ruleres} on $x_2$};
\node[sys] (b) at (2.4, -1.6) {$e(\overline{x_1})$\\ \scriptsize {\rulecl} $1 \leq \overline{x_1}$};
\node[sys] (c) at (-4.6, -3.4) {$e(x_1) + e(x_2)$\\ \scriptsize {\rulecl} $1 \leq x_1 \odot x_2$};
\node[sys] (d) at (-0.4, -3.4) {$e(x_1) + e(\overline{x_2})$\\ \scriptsize {\rulecl} $1 \leq x_1 \odot \overline{x_2}$};
\draw[sim] (r) -- (a) node[lab, midway, above, sloped] {$0$};
\draw[sim] (r) to[bend left=18] node[lab, midway, above, sloped] {$0$} (b);
\draw[sim] (b) to[bend left=18] node[lab, midway, below, sloped] {$1$} (r);
\draw[sim] (a) -- (c) node[lab, midway, above, sloped] {$0$};
\draw[sim] (a) -- (d) node[lab, midway, above, sloped] {$0$};
\draw[sim] (c) .. controls (-5.6, -1.2) and (-3.8, 0.9) .. node[lab, pos=0.45, above, sloped] {$1$} (r);
\draw[sim] (d) to[bend right=20] node[lab, midway, below, sloped] {$1$} (r);
\end{tikzpicture}
\caption{A refutation strategy graph.}
\label{fig:toy-subgraph}
\end{subfigure}

\medskip
\begin{subfigure}{\textwidth}
\centering
\begin{tikzpicture}[x=1cm, y=1cm]
\node[sys] (st) at (0, 0) {$0 \leq 0$\\ \scriptsize {\ruleres} on $x_1$};
\node[sys] (u1) at (-2.4, -1.6) {$x_1 \leq 0$\\ \scriptsize {\ruleres} on $x_2$};
\node[sys] (u2) at (2.4, -1.6) {$\overline{x_1} \leq 0$\\ \scriptsize {\rulecl}};
\node[sys] (v1) at (-4.0, -3.2) {$x_1 + x_2 \leq 0$\\ \scriptsize {\rulecl}};
\node[sys] (v2) at (-0.8, -3.2) {$x_1 + \overline{x_2} \leq 0$\\ \scriptsize {\rulecl}};
\node[sys] (t) at (0, -4.8) {$0 \leq -1$};
\draw[sim] (st) -- (u1);
\draw[sim] (st) -- (u2);
\draw[sim] (u1) -- (v1);
\draw[sim] (u1) -- (v2);
\draw[sim] (v1) -- (t);
\draw[sim] (v2) -- (t);
\draw[sim] (u2) -- (t);
\end{tikzpicture}
\caption{A truncated unfolding: a backward proof.}
\label{fig:toy-backward}
\end{subfigure}
\caption{A refutation strategy graph for $F = (x_1 \vee x_2) \wedge (x_1 \vee \neg x_2) \wedge \neg x_1$ and a backward {\thrres} proof obtained by unfolding its cycles (Example~\ref{ex:toy-strategy}). For clarity, the illustrated unfolding stops when $0\leq-1$ is first reached and identifies the terminal copies; it is a truncation of the general construction in Section~\ref{sec:unfolding}, which continues to its boundary sinks. At the top, each vertex is shown with its position and type, and the edge labels are the weights; at the bottom, each vertex is shown with its threshold label and type.}
\label{fig:toy-strategy}
\end{figure}
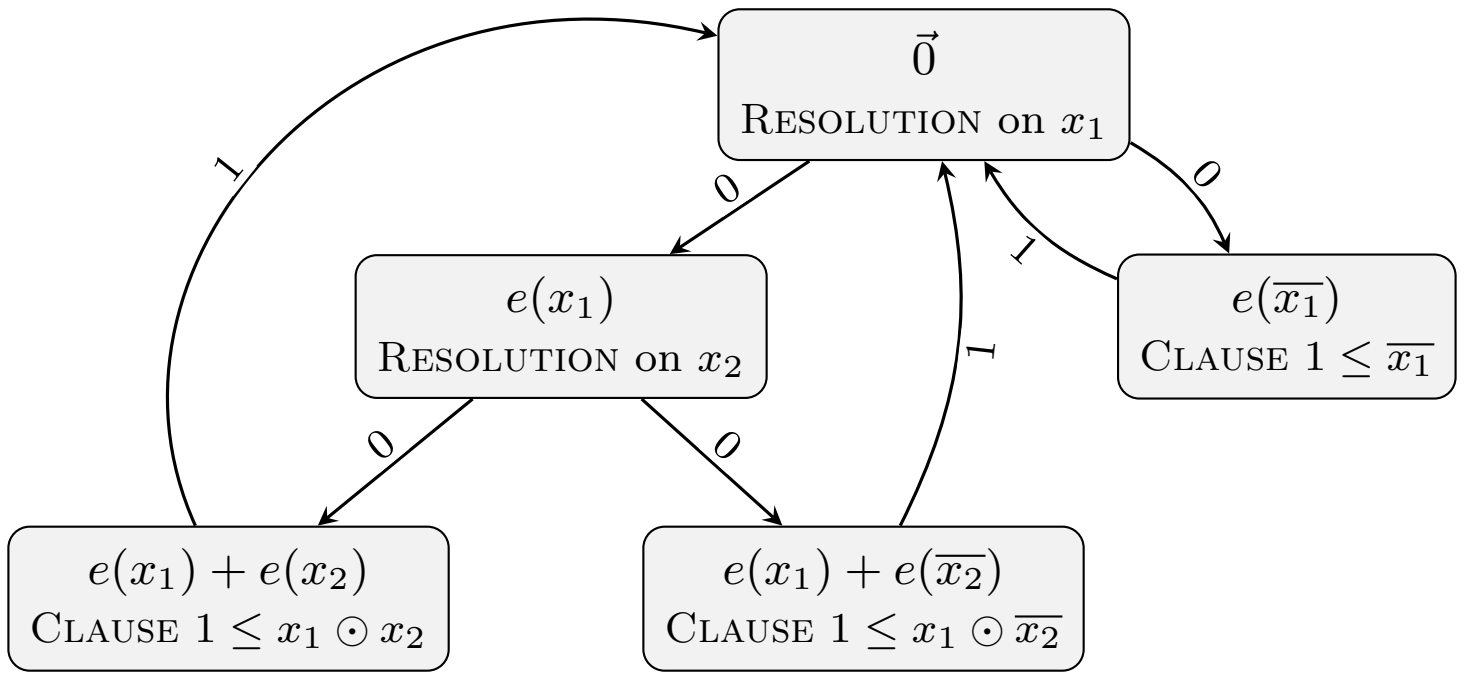
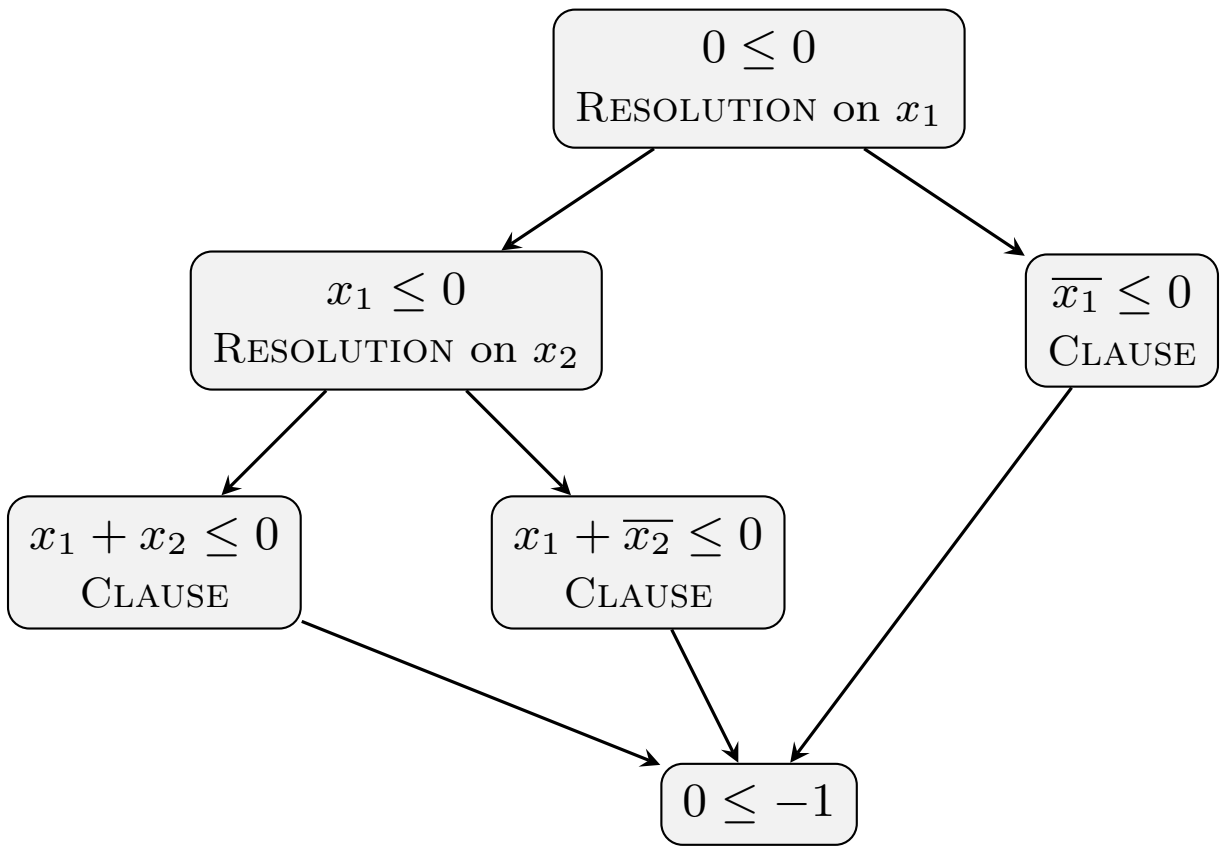

The proof uses the following two ideas. First, a static {\mpnsr} certificate determines a finite weighted \emph{strategy graph} on monomial positions: at every monomial position the certificate selects one premise to apply, the outgoing edges of the position describe the obligations resulting from that premise, and the global consistency of the certificate appears as the condition that every directed cycle of the graph has strictly positive total weight. Second, a finite graph with strictly positive cycles and local threshold moves at the vertices can be \emph{unfolded} into a finite acyclic proof: tracking the accumulated weight explicitly turns each vertex into polynomially many copies, and the positive-cycle condition guarantees that every branch of the unfolded proof terminates. The converse direction compresses a daglike threshold proof back into such a graph by selecting, at every monomial position, an occurrence with the maximum threshold. The proof proceeds through the following chain of translations
\begin{gather*}
\text{{\mpnsr} refutation}
\;\overset{\text{T\ref{thm:refut-subgraph-equiv}}}{\longleftrightarrow}\;
\text{refutation subgraph}
\;\overset{\text{L\ref{lm:normal-form}}}{\longleftrightarrow}\;
\text{strategy graph}\\
\;
\xleftrightarrow{\text{L\ref{lm:unfolding-lemma},\ P\ref{prop:compression}}}\;
\text{backward proof}
\;\overset{\text{P\ref{prop:reversal}}}{\longleftrightarrow}\;
\text{{\thrres} refutation.}
\end{gather*}
Section~\ref{sec:mp-graph} states the graph-theoretic characterization of {\mpns} refutations (the first double arrow). Section~\ref{sec:contraction} contracts the resulting subgraphs into \emph{strategy graphs} with a short list of local transition types. Section~\ref{sec:backward} introduces the backward reading of Threshold Resolution proofs, and Section~\ref{sec:unfolding} unfolds a strategy graph into a backward proof; Section~\ref{sec:compression} proves the converse by compressing a backward proof into a strategy graph. Section~\ref{sec:assembly} assembles the pieces. Figure~\ref{fig:toy-strategy} shows the two middle objects of the chain --- a strategy graph and its unfolding into a backward proof --- for a toy formula.

Throughout this section we fix an unsatisfiable CNF $F$ over the variables $x_1, \ldots, x_N$ and work with the {\mpnsr} encoding of $F$ (Definitions~\ref{def:mpnsr} and~\ref{def:clause-encoding}): the premises of the refuted system are the clause encodings~\eqref{eq:encoding-i}, the standard axioms and the resolution axioms. Monomials in the dual-encoded variables are identified with their exponent vectors: a \emph{position} is a vector $v \in \Z_+^{2N}$ with a coordinate $v_l$ for every literal $l$; we write $e(l)$ for the unit vector of the literal $l$, and
\[
f_v \;:=\; \sum_{l} v_l\, l
\]
for the linear form (equivalently, the multiset of literals) at the position $v$, so that $|f_v| = \sum_l v_l$. The origin $\vec 0$ is the position of the constant monomial.

\subsection{Monomial certificates and finite refutation graphs}
\label{sec:mp-graph}

Consider an {\mpns} refutation: at every monomial $m$ of the left-hand side of the resulting combination, the coefficient of $m$ on the left must strictly exceed the coefficient of $m$ on the right (Definition~\ref{def:contradictory}). Both coefficients are minima of finitely many \emph{occurrences}: an entry of the certificate using a premise $f_i \leq g_i$ with the term $\lambda \odot s$ places the occurrence $\lambda + \coef_q(g_i)$ at the monomial $s \odot q$ for every monomial $q$ of $g_i$, and symmetrically with $f_i$ on the left --- an occurrence is a coefficient of a premise, shifted by a multiplier of the certificate. The condition at $m$ therefore alternates a choice with a requirement: \emph{at least one} right-hand occurrence at $m$ must be strictly smaller than \emph{every} left-hand occurrence at $m$. This alternation is what generates a bipartite graph with an OR side and an AND side. At a monomial it suffices to select one witnessing right-hand occurrence (OR); the selected occurrence names a premise and a shift, and the left-hand side of that premise places left-hand occurrences at all the monomials it touches --- the condition must hold at each of them, so all of them become new obligations (AND). Iterating from the constant monomial, whose right-hand coefficient is finite by Definition~\ref{def:contradictory}, produces a finite bipartite graph, in which the monomials form the OR side and the premise occurrences the AND side, and the strict per-monomial inequalities, telescoped along any directed cycle, force every directed cycle to have strictly positive total weight. The construction below generalizes the coefficient graphs used by Alekseev, Grigoriev and Hirsch in their lower-bound arguments for tropical binomials, where a selected occurrence on the right-hand side determines an edge to a monomial with a strictly smaller coefficient (\cite{AGH25} Section~7.5, Theorems~7.15--7.16).

\begin{definition}[Monomial graph]
\label{def:monomial-graph}
Given a set of tropical inequalities (premises) $\{f_i \leq g_i : i \in [M]\}$ over $n$ variables with coefficients in $\Q_\infty$, the \emph{monomial graph} is the following infinite bipartite directed graph. The left-hand side vertices are the pairs $(s, i)$, where $s$ is a monomial over the $n$ variables and $i \in [M]$; the right-hand side vertices are the monomials. The edges are:
\begin{enumerate}
\item for each $i \in [M]$, each monomial $s$, and each monomial $t$ occurring in $f_i$: an edge from $(s, i)$ to the right vertex $s \odot t$, of weight $\coef_t(f_i)$;
\item for each $i \in [M]$, each monomial $m$, and each monomial $q$ occurring in $g_i$ that divides $m$: an edge from the right vertex $m$ to $(m / q,\, i)$, of weight $-\coef_q(g_i)$.
\end{enumerate}
\end{definition}

\begin{definition}[Refutation subgraph]
\label{def:refutation-subgraph}
A subgraph $R$ of the monomial graph is a \emph{refutation subgraph} if:
\begin{enumerate}
\item $R$ has finitely many vertices;
\item $R$ contains the right vertex $\vec 0$ (the constant monomial);
\item every vertex of $R$ is reachable from $\vec 0$;
\item every right vertex of $R$ keeps exactly one of its outgoing edges;
\item every left vertex of $R$ keeps all of its outgoing edges;
\item every directed cycle of $R$ has strictly positive total weight.
\end{enumerate}
Keeping an edge includes keeping both of its endpoints.
\end{definition}

\begin{remark}
\label{rem:mean-payoff}
This one-choice/all-responses structure is analogous to the positional strategies arising in the equivalence between tropical feasibility and mean-payoff games (\cite{AGG12} Theorem~2.13 and Section~3). The characterization below is nevertheless proved directly: the vertices here encode monomial shifts, and we need explicit proof-size and degree bounds.
\end{remark}

The reconstruction of a certificate from a subgraph rests on the following graph fact, a version of the Johnson potentials~\cite{Joh77} from shortest-path reweighting.

\begin{lemma}[Positive-cycle potentials]
\label{lm:potentials}
Let $R$ be a finite rooted bipartite weighted digraph, with every vertex reachable from the root, in which every directed cycle has strictly positive total weight. Then its vertices can be assigned rational values, denoted by $\lambda_{s,i}$ at the left vertices and $\rho(m)$ at the right vertices, such that
\[
\lambda_{s,i} + a \;>\; \rho(m)
\qquad \text{for every left-to-right edge } (s,i) \to m \text{ of weight } a,
\]
and
\[
\rho(m) + b \;\geq\; \lambda_{s,i}
\qquad \text{for every right-to-left edge } m \to (s,i) \text{ of weight } b.
\]
If all edge weights are integers of absolute value at most $W$ and $r$ is the number of right vertices, the values can be chosen with the common denominator $r + 1$ and unary size $O(|R|^2 (W+1))$ each; for rational edge weights of maximum bitsize $B$, they can be chosen of bitsize $\poly(|R|, B)$.
\end{lemma}
\begin{proof}
Assume first that the weights are integers. Subtract $\tfrac{1}{r+1}$ from the weight of every left-to-right edge and leave the right-to-left weights unchanged. Every simple directed cycle is alternating, so it contains at most $r$ left-to-right edges, and its original weight is a positive integer, hence at least $1$; its perturbed weight is therefore at least $1 - \tfrac{r}{r+1} > 0$, and the perturbed graph has no cycle of negative (or zero) weight --- in particular, this covers the acyclic case. Since every vertex is reachable from the root and no negative cycles exist, the shortest-path distance $h(v)$ from the root with respect to the perturbed weights $w'$ is well defined and finite, and satisfies $h(v) \leq h(u) + w'(u, v)$ for every edge $(u, v)$. Set $\lambda_{s,i} := h((s,i))$ and $\rho(m) := h(m)$. For a left-to-right edge of weight $a$ this gives $\rho(m) \leq \lambda_{s,i} + a - \tfrac{1}{r+1} < \lambda_{s,i} + a$, the required strict inequality, and for a right-to-left edge of weight $b$ it gives $\lambda_{s,i} \leq \rho(m) + b$. Each value is the weight of a path with fewer than $|R|$ edges, hence a rational with denominator dividing $r + 1$ and absolute value at most $|R|(W + 1)$, of unary size $O(|R|^2(W+1))$.

For rational weights, multiply all weights by a common denominator --- its bitsize is polynomial in the total bitsize of the weights --- apply the integral case, and scale the resulting values back.
\end{proof}

\begin{theorem}[Certificate--graph characterization]
\label{thm:refut-subgraph-equiv}
Let $\mathcal{S} = \{f_i \leq g_i : i \in [M]\}$ be a finite set of tropical inequalities; let $L$ bound the number of monomials, $d$ the degrees of the monomials, and $B$ the bitsizes of the coefficients occurring in a single premise.
\begin{enumerate}
\item An {\mpns} refutation of $\mathcal{S}$ of size $S$ yields a refutation subgraph of the monomial graph with at most $S(L + 1)$ vertices whose labels are monomials of degree at most $S(L+1)d$.
\item A refutation subgraph with $r$ vertices and labels of degree at most $D$ yields an {\mpns} refutation of $\mathcal{S}$ of size $\poly(r, D, B)$. If the coefficients of the premises are integers of absolute value at most $W$, the multipliers of the refutation can moreover be chosen with a common denominator at most $r + 1$ and unary size $\poly(r, W)$.
\end{enumerate}
\end{theorem}
\begin{proof}
\emph{From certificates to subgraphs.} By Remark~\ref{rem:trivial-static}, delete all trivial-axiom entries without increasing the size, so every remaining entry uses a premise of $\mathcal{S}$. Write each term of the refutation as $t_j = \lambda_j \odot s_j$ and let $P \leq Q$ be the combination~\eqref{eq:mp-ineq}, so that $P \succ Q$. First normalize the certificate: if two entries use the same premise $f_i \leq g_i$ with the same monomial shift $s$, retain one of them with the smaller coefficient --- this changes neither $P$ nor $Q$ and does not increase the size --- so that one multiplier $\lambda_{s,i} \odot s$ is associated with each used pair $(s, i)$.

For every monomial $m$ occurring in $Q$, set $\rho(m) := \coef_m(Q)$ and choose a premise pair $(s, i)$ and a monomial $q$ occurring in $g_i$ that attain this coefficient:
\[
m = s \odot q, \qquad \rho(m) = \lambda_{s,i} + \coef_q(g_i);
\]
such a choice exists, since $\coef_m(Q)$ is the attained minimum of the finitely many contributions $\lambda_{s,i} + \coef_q(g_i)$ with $s \odot q = m$. Keep the right-to-left edge $m \to (s, i)$ given by $q$, keep every outgoing edge of every selected left vertex, and retain the part reachable from $\vec 0$; the root occurs in $Q$ since $\coef_{\vec 0}(Q)$ is finite. The selection is closed: if an edge $(s, i) \to m' = s \odot t$ is kept, then $\coef_{m'}(P) \leq \lambda_{s,i} + \coef_t(f_i) < \infty$, so $m'$ occurs in $P$, hence also in $Q$ because $P \succ Q$, and an outgoing edge is selected at $m'$. The result satisfies conditions 1--5 of Definition~\ref{def:refutation-subgraph}: it is finite (its right vertices occur in $Q$ and its left vertices are used pairs), rooted, every right vertex keeps one edge, and every left vertex keeps all of them.

For the cycles, consider a kept two-step path $m \to (s, i) \to m' = s \odot t$, of total weight $\coef_t(f_i) - \coef_q(g_i)$. The contradiction condition gives
\[
\rho(m') \;=\; \coef_{m'}(Q) \;<\; \coef_{m'}(P) \;\leq\; \lambda_{s,i} + \coef_t(f_i) \;=\; \rho(m) + \coef_t(f_i) - \coef_q(g_i).
\]
Every directed cycle decomposes into such two-step paths, and summing the displayed inequality around the cycle cancels the values $\rho$: the total weight is strictly positive, which is condition~6 (the total weight of the cycle is exactly the sum of the values $\coef_t(f_i) - \coef_q(g_i)$ over its two-step paths).

The counts: the left vertices are among the $K \leq S$ normalized pairs and the right vertices occur in $Q$, so there are at most $S(L+1)$ vertices. Every vertex is reachable from $\vec 0$ by a simple path, and each edge changes the degree of its monomial label by at most $d$ (the label of a left vertex $(s,i)$ is $s$). Thus all labels have degree at most $S(L+1)d$.

\emph{From subgraphs to certificates.} Let $R$ be a refutation subgraph with $r$ vertices; its edge weights are coefficients of the premises. Apply Lemma~\ref{lm:potentials} to $R$, obtaining values $\lambda_{s,i}$ and $\rho(m)$. Take the certificate that uses, for every left vertex $(s, i)$ of $R$, the premise $f_i \leq g_i$ with the term $\lambda_{s,i} \odot s$, and nothing else; let $P \leq Q$ be the resulting combination.

Let $m$ occur in $P$. Every contribution to $\coef_m(P)$ comes from a kept edge $(s, i) \to m$ of weight $a = \coef_t(f_i)$, since left vertices keep all their edges; the potential inequality $\lambda_{s,i} + a > \rho(m)$ for each of them gives $\coef_m(P) > \rho(m)$. Moreover, $m$ is then a right vertex of $R$ and has its selected edge $m \to (s', i')$, given by a monomial $q$ with $m = s' \odot q$ and weight $-\coef_q(g_{i'})$; the other potential inequality, $\rho(m) - \coef_q(g_{i'}) \geq \lambda_{s',i'}$, shows that the corresponding contribution to $\coef_m(Q)$ is at most $\rho(m)$, and further right-hand-side contributions can only decrease a tropical coefficient. Hence
\[
\coef_m(Q) \;\leq\; \lambda_{s',i'} + \coef_q(g_{i'}) \;\leq\; \rho(m) \;<\; \coef_m(P).
\]
Finally, the root $\vec 0$ is a right vertex of $R$, so $\coef_{\vec 0}(Q) \leq \rho(\vec 0) < \infty$. Hence $P \succ Q$: the certificate is an {\mpns} refutation.

Its size: there are at most $r$ entries, the monomial shifts are labels of $R$ and have degree at most $D$, and the multipliers are the potentials of Lemma~\ref{lm:potentials}, of bitsize $\poly(r, B)$ --- in total $\poly(r, D, B)$. For integer premise coefficients of absolute value at most $W$, the integral case of the lemma gives multipliers with a common denominator at most $r + 1$ and unary size $\poly(r, W)$.
\end{proof}

\begin{corollary}[Unary certificates]
\label{cor:mpnsr-degree}
\label{cor:unary-coefficients}
Every binary {\mpnsr} refutation of size $S$ of the encoding of a CNF $F$ on $N$ variables can be transformed into a refutation of size $\poly(S,N)$ with unary exponents, maximum degree $\poly(S,N)$, and rational coefficients of unary size $\poly(S,N)$ with a common denominator $\poly(S,N)$.
\end{corollary}
\begin{proof}
For the standard {\mpnsr} encoding, every premise has at most three monomials in total, of degree at most $D_0:=\max\{2,w\}$, where $w\leq2N$ is the maximum clause width. Theorem~\ref{thm:refut-subgraph-equiv}, with $L\leq3$ and $d\leq D_0$, gives a refutation subgraph with $r\leq4S$ vertices and labels of degree at most $4SD_0=\poly(S,N)$. The premise coefficients are $0$ and $\pm1$, so the edge weights are integers with $W=1$. The converse direction of the theorem therefore reconstructs a certificate with these monomial labels and coefficients of unary size $\poly(r,1)$ with a common denominator at most $r+1$. Writing the exponents in unary gives the required refutation of size $\poly(S,N)$.
\end{proof}

\begin{remark}[Beyond CNFs]
\label{rem:unary-beyond-cnf}
The proof of Corollary~\ref{cor:unary-coefficients} uses the CNF only to bound $L$, $d$ and the premise coefficients: for any system of tropical inequalities with integer coefficients of absolute value at most $W$, it turns a binary {\mpnsr} refutation of size $S$ into a unary one of size $\poly(S, N, L, d, W)$. For premises with exponentially large numbers written in binary, unary refutations can be exponentially longer. For example, multiplying the premises $2^n \odot x_1 \leq 2^n$ and $1 \leq x_1$ by $-2^n$ and $-\tfrac12$, respectively, gives a binary {\mpnsr} refutation of size $O(n)$, but every {\mpnsr} refutation of these premises has a multiplier of absolute value at least $2^{n-2}$. Indeed, by Remark~\ref{rem:trivial-static} we may assume that no trivial axioms are used. If all multipliers were smaller, every coefficient contributed by the first premise would exceed every coefficient contributed by the other entries; deleting the entries of the first premise would then keep $\coef_m(P) > \coef_m(Q)$ for every monomial $m$ of the left-hand side $P$ (some entries remain, since the first premise together with the axioms is satisfied by $x_1 = 0$). This is impossible, since all remaining premises hold at $x_1 = 1$, $\overline{x_1} = 0$, where the left-hand side would then be strictly greater than the right-hand side.
\end{remark}

\subsection{Contraction: from refutation subgraphs to strategy graphs}
\label{sec:contraction}

In a refutation subgraph every right vertex selects one premise, and every left vertex is passed through in one way; contracting each length-two path
\[
\text{monomial} \longrightarrow \text{premise occurrence} \longrightarrow \text{monomial}
\]
into a single weighted edge leaves a graph on monomial positions alone, with one \emph{local transition type} per possible selected premise. We first name the structural properties of the contracted graphs.

\begin{definition}[Strategy graph]
\label{def:strategy-graph}
A \emph{strategy graph} is a finite directed graph whose vertices are positions $v \in \Z_+^{2N}$, rooted at the origin $\vec 0$, in which every vertex is reachable from the root, every vertex has out-degree at least $1$, all edge weights are integers, and every directed cycle has strictly positive total weight.

A strategy graph is a \emph{refutation strategy graph for $F$} if every vertex, together with its outgoing edges, is of one of the five local types of Table~\ref{tab:local-types}. All edge weights of a refutation strategy graph lie in $\{-1, 0, 1\}$.
\end{definition}

\begin{table}[ht]
\centering
\footnotesize
\setlength{\tabcolsep}{3pt}
\begin{tabular}{llll}
type & applied premise & edges at the position $v$ & backward move at $f_v \leq t$\\
\hline
{\rulecl} & $1 \leq \bigodot_{l \in C} l$, clause $C$ of $F$; & edge to $v - \sum_{l \in C} e(l)$, & child $f_v - \sum_{l \in C} l \leq t - 1$\\
 & requires $v \geq \sum_{l \in C} e(l)$ & weight $1$ & \\[2pt]
{\ruleres} & $x_i \oplus \overline{x_i} \leq 0$ & edges to $v + e(x_i)$ and & children $f_v + x_i \leq t$\\
 & & to $v + e(\overline{x_i})$, weight $0$ & and $f_v + \overline{x_i} \leq t$\\[2pt]
{\ruleweak} & $0 \leq l$; requires $v_l \geq 1$ & edge to $v - e(l)$, weight $0$ & child $f_v - l \leq t$\\[2pt]
{\ruleintro} & $1 \leq x_i \odot \overline{x_i}$; requires & edge to $v - e(x_i) - e(\overline{x_i})$, & child $f_v - x_i - \overline{x_i} \leq t - 1$\\
 & $v_{x_i}, v_{\overline{x_i}} \geq 1$ & weight $1$ & \\[2pt]
{\rulesim} & $x_i \odot \overline{x_i} \leq 1$ & edge to $v + e(x_i) + e(\overline{x_i})$, & child $f_v + x_i + \overline{x_i} \leq t + 1$\\
 & & weight $-1$ & \\
\end{tabular}
\caption{The local transition types of a refutation strategy graph for $F$. The last column gives the backward Threshold Resolution reading of Section~\ref{sec:backward}: a vertex at the position $v$ carries a threshold label $f_v \leq t$, and an edge of weight $w$ leads to a child whose label has the threshold $t - w$.}
\label{tab:local-types}
\end{table}

\begin{lemma}[Normal form]
\label{lm:normal-form}
A refutation subgraph of the monomial graph of the {\mpnsr} encoding of $F$ exists if and only if a refutation strategy graph for $F$ exists, and their sizes are polynomially related.
\end{lemma}
\begin{proof}
We first compute the local structure of the monomial graph at a left vertex. Every premise of the encoding has a single term on the right-hand side; write it as $\min_{j} (a_j^T x - b_j) \leq c^T x - d$ with $a_j, c \in \Z_+^{2N}$, $b_j, d \in \Z$. By Definition~\ref{def:monomial-graph}, the left vertex $(s, i)$ receives its incoming edges from the right vertex $s \odot x^c$, with weight $-\coef_c(g_i) = d$, and sends one outgoing edge to $s \odot x^{a_j}$, of weight $\coef_{a_j}(f_i) = -b_j$, for every $j$.

Let $R$ be a refutation subgraph. Every right vertex of $R$ keeps exactly one outgoing edge, so we may replace each kept length-two path $m \to (s, i) \to m'$ by the single edge $m \to m'$ whose weight is the sum of the two weights; the contracted graph is rooted at $\vec 0$, since the paths from $\vec 0$ in $R$ alternate between the two sides and correspond to contracted paths. Instantiating the computation above for each premise of the encoding shows that a right vertex $m$ acquires exactly the edges of the type of Table~\ref{tab:local-types} named by its selected premise: for the resolution axiom $x_i \oplus \overline{x_i} \leq 0$ the incoming edge of the left vertex comes from $m$ itself with weight $0$ and the two outgoing edges lead to $m + e(x_i)$ and $m + e(\overline{x_i})$ with weights $0$; for a clause encoding $1 \leq \bigodot_{l \in C} l$ the incoming edge comes from $m$ with $m \geq \sum_{l \in C} e(l)$, and the single outgoing edge leads to $m - \sum_{l \in C} e(l)$ with total weight $0 + 1$; the standard axioms are checked in the same way. Every cycle of the contracted graph corresponds bijectively to a cycle of $R$ of the same total weight, hence all cycles are strictly positive; every vertex keeps its single contracted edge or edges, so the out-degree is at least $1$. The result is a refutation strategy graph for $F$ of at most the size of $R$.

Conversely, given a refutation strategy graph $Q$, select for each vertex the premise named by its type and, in the monomial graph, the corresponding right-to-left edge together with all outgoing edges of the chosen left vertex. The edges of $Q$ correspond to the resulting length-two paths with the same weights, so cycles keep their weights, reachability from $\vec 0$ is inherited along the expanded paths, and the result is a refutation subgraph of size polynomial in the size of $Q$.
\end{proof}

\begin{example}
\label{ex:toy-strategy}
Let $F = (x_1 \vee x_2) \wedge (x_1 \vee \neg x_2) \wedge \neg x_1$, with the clauses encoded by $1 \leq x_1 \odot x_2$, $1 \leq x_1 \odot \overline{x_2}$ and $1 \leq \overline{x_1}$. Figure~\ref{fig:toy-strategy} (top) shows a refutation strategy graph for $F$: the origin and the position $e(x_1)$ are {\ruleres} vertices, on $x_1$ and on $x_2$ respectively, and the positions $e(\overline{x_1})$, $e(x_1) + e(x_2)$ and $e(x_1) + e(\overline{x_2})$ are {\rulecl} vertices for the three clauses of $F$. The graph has three directed cycles through the origin, each of total weight $1 > 0$. Figure~\ref{fig:toy-strategy} (bottom) shows the truncated unfolding described in the caption: tracking the thresholds explicitly cuts the cycles, and the copy of the origin at threshold $-1$, reached along all three cycles, carries the label $0 \leq -1$: the terminal. Reversed (Proposition~\ref{prop:reversal}), it is the six-line {\thrres} refutation that introduces the three clause encodings by {\rulecl} steps, resolves $x_1 + x_2 \geq 1$ and $x_1 + \overline{x_2} \geq 1$ on $x_2$ into $x_1 \geq 1$, and resolves $x_1 \geq 1$ against $\overline{x_1} \geq 1$ into $0 \geq 1$.
\end{example}

\subsection{The backward presentation of Threshold Resolution}
\label{sec:backward}

For the conversion between graphs and proofs it is convenient to read a {\thrres} proof backwards, from the contradiction towards the source. Reverse every proof edge and replace each line $f \geq s+1$ by the condition $f \leq s$, its pointwise negation over the Boolean points. The source $0 \geq 0$ becomes the terminal condition $0 \leq -1$, while the contradiction $0 \geq 1$ becomes the start condition $0 \leq 0$. A \emph{backward proof} is thus a finite DAG of threshold conditions $f \leq t$ with a start vertex labeled $0 \leq 0$ and a single sink, the \emph{terminal}, labeled $0 \leq -1$, in which every non-terminal vertex, together with its outgoing edges, performs one of the five moves in the last column of Table~\ref{tab:local-types} (in an arbitrary context: the labels shown may be shifted by a common multiset of literals and a common threshold).

A backward proof is \emph{duplicate-free} if no two vertices carry the same condition $f\leq t$. Reversal preserves this property, so Remark~\ref{rem:no-duplicates} also allows us to remove duplicate lines from a backward proof without increasing its size.

\begin{proposition}[Reversal]
\label{prop:reversal}
The reversal above is an involution between {\thrres} refutations, trimmed to the ancestors of the line $0 \geq 1$, and backward proofs: it preserves size and turns every rule application into the backward move of the same row of Table~\ref{tab:local-types}, and vice versa.
\end{proposition}
\begin{proof}
Direct check of the conditions above.
\end{proof}

Backward proofs will be produced by the unfolding below, which also needs a supply of moves for entering the terminal from any label that is violated by every Boolean point outright:

\begin{lemma}[Cleanup]
\label{lm:cleanup}
For every position $v$ and every integer $t \leq -2$ there is a chain of $O(|f_v| + |t|)$ backward moves of the types {\ruleweak} and {\rulesim} leading from a vertex labeled $f_v \leq t$ to the terminal $0 \leq -1$.
\end{lemma}
\begin{proof}
{\ruleweak} moves strip the literals of $f_v$ one at a time (a vertex $f + l \leq t$ has the child $f \leq t$), reaching the label $0 \leq t$ in $|f_v|$ moves. Then repeat the following round $|t| - 1$ times, raising the threshold by one: a {\rulesim} move from $0 \leq u$ to $x_1 + \overline{x_1} \leq u + 1$, followed by two {\ruleweak} moves removing the introduced pair, ends at $0 \leq u + 1$. The chain ends at $0 \leq -1$.
\end{proof}

\subsection{Unfolding: from strategy graphs to backward proofs}
\label{sec:unfolding}

Two things separate a refutation strategy graph from a backward proof. First, the graph may contain directed cycles --- necessarily of positive weight --- while a proof is acyclic. Second, only the differences of the thresholds are recorded (the edge weights), not the thresholds themselves, and nothing guarantees that every branch reaches the terminal. Both are resolved by one combinatorial construction: layer the graph by the accumulated weight. We state it as a purely graph-theoretic lemma, independent of the proof-system semantics; the meanings of vertices and labels enter only in the Unfolding Lemma that follows.

\begin{lemma}[Positive-cycle layering]
\label{lm:layering}
Let $G = (V, E, w)$ be a finite rooted directed graph with root $r$ in which every vertex is reachable from $r$, every vertex has positive out-degree, and the edge weights are integers with $|w(e)| \leq W'$ for all $e \in E$, where $W' \geq 1$. Suppose that every directed cycle of $G$ has strictly positive total weight. Then there exist a finite rooted DAG $G'$, a projection $\pi : V(G') \to V$ and an integer potential $\tau : V(G') \to \Z$ such that:
\begin{enumerate}
\item The root $r'$ of $G'$ satisfies $\pi(r') = r$ and $\tau(r') = 0$.
\item If $u'$ is not a sink of $G'$, then its outgoing edges are in bijection with the outgoing edges of $\pi(u')$ in $G$.
\item If an edge out of $u'$ corresponds to an edge $e = (\pi(u'), v)$ of $G$, then its endpoint $v'$ satisfies $\pi(v') = v$ and $\tau(v') = \tau(u') - w(e)$.
\item Every sink $u'$ of $G'$ satisfies $\tau(u') \leq -2$.
\item Writing $n = |V|$ and $m = |E|$, the graph $G'$ has $O(n^2 W')$ vertices and $O(n m W')$ edges.
\end{enumerate}
\end{lemma}
\begin{proof}
Let $n := |V|$ and $H := (n+1)W' + 1$, and consider the set of \emph{layered states} $V \times [-H, H]_{\Z}$; the intended meaning of a state $(v, t)$ is that the original graph is at $v$ while $t$ records the negative of the accumulated edge weight. Call a state $(v, t)$ \emph{complete} if $t - w(e) \in [-H, H]$ for every outgoing edge $e$ of $v$ in $G$. For every complete state $(v, t)$ and every outgoing edge $e = (v, u)$, add the layered edge $(v, t) \to (u, t - w(e))$; the remaining states carry no outgoing edges. Let $G'$ be the part of the layered graph reachable from $(r, 0)$, rooted there, and set $\pi(v, t) := v$ and $\tau(v, t) := t$. Properties~1--3 hold by construction: every non-sink is complete and carries exactly one layered edge per outgoing edge of its projection.

\emph{Acyclicity.} A directed cycle of $G'$ would project to a closed directed walk in $G$; along every layered edge $t_{i+1} = t_i - w(v_i, v_{i+1})$, and the potential returns to its initial value around the cycle, so the projected walk has total weight $0$. But every nonempty closed directed walk decomposes into directed cycles, each of strictly positive integer weight, so its total weight is at least $1$ --- a contradiction. Hence $G'$ is acyclic.

\emph{Upper bound on reachable potentials.} Let $(v, t)$ be reachable from $(r, 0)$ and project a path from $(r, 0)$ to $(v, t)$ onto a directed walk in $G$; by the definition of the layered edges its total weight is $-t$. Removing directed cycles --- each of weight at least $1$ --- leaves a simple path with fewer than $n$ edges, of weight at least $-nW'$. Consequently $-t \geq -nW'$, that is,
\begin{equation}
\label{eq:potential-bound}
t \;\leq\; n W' .
\end{equation}

\emph{Sinks.} Since every vertex of $G$ has positive out-degree, a state $(v, t)$ of $G'$ is a sink exactly when some outgoing edge weight $w$ of its projection has $t - w \notin [-H, H]$. The upper boundary is unreachable: by~\eqref{eq:potential-bound}, $t - w \leq nW' + W' = (n+1)W' < H$. Hence $t - w < -H$, and using $w \leq W'$ we get $t < -H + W' = -nW' - 1$; since $n, W' \geq 1$ and $t$ is an integer, $t \leq -2$. This proves Property~4.

\emph{Size.} There are at most $n(2H + 1) = O(n^2 W')$ layered states, and each original edge has at most one lifted copy per potential level, so at most $m(2H+1) = O(nmW')$ layered edges; restricting to the reachable part only decreases these counts.
\end{proof}

\begin{lemma}[Unfolding]
\label{lm:unfolding-lemma}
Let $G$ be a refutation strategy graph for $F$ with $r$ vertices, edge weights of absolute value at most $W'$, and positions of $\ell_1$-norm at most $P$. Then $F$ has a backward {\thrres} proof --- hence, by Proposition~\ref{prop:reversal}, a {\thrres} refutation --- of size $\poly(r, W', P)$, with all thresholds of absolute value $O(rW')$.
\end{lemma}
\begin{proof}
A strategy graph satisfies the hypotheses of Lemma~\ref{lm:layering}; let $G'$, $\pi$, $\tau$ be as provided by it, and label every state $u'$ of $G'$ by the threshold condition $f_{\pi(u')} \leq \tau(u')$. The root is labeled $f_{\vec 0} \leq 0$, the start condition $0 \leq 0$.

Every non-sink state $(v, t)$ keeps, by Properties~2--3, one outgoing edge per edge of $v$ in $G$, and the child along an edge of weight $w$ to $u$ is labeled $f_u \leq t - w$. Comparing with Table~\ref{tab:local-types}: the vertex $v$ is of one of the five local types, and the labels of $(v, t)$ and of its children are exactly the vertex label and the child labels of the backward move of that type at $f_v \leq t$. So every non-sink of $G'$ is a valid backward-move vertex.

Every sink $u'$ satisfies $\tau(u') \leq -2$ by Property~4, so Lemma~\ref{lm:cleanup} provides a chain of $O(P + rW')$ backward moves from its label to the terminal $0 \leq -1$; attach a fresh copy of this chain below every sink, and let all chains share one terminal vertex.

The result is a backward proof: the start is the root; the graph is acyclic, since $G'$ is a DAG and the chains are fresh vertices attached forward; every non-terminal vertex performs a backward move of Table~\ref{tab:local-types}; and every maximal path enters some sink of $G'$ and then runs down its cleanup chain to the terminal --- the single sink of the whole proof.

The size: $G'$ has $O(r^2 W')$ states and $O(r \cdot |E| \cdot W')$ edges with $|E| \leq 2r$ (each vertex has at most two outgoing edges), each label has size $O(P + rW')$ by~\eqref{eq:potential-bound} and Property 4 combined with $|\tau| \leq H + 1$, and each of the $O(r^2W')$ possible chains has length $O(P + rW')$; the total is $\poly(r, W', P)$, and all thresholds are bounded in absolute value by $H + 1 = O(rW')$.
\end{proof}

\subsection{Compression: from backward proofs to strategy graphs}
\label{sec:compression}

The converse direction compresses a backward proof into a finite strategy graph. The proof may visit the same position many times, with different thresholds, and the graph has room for only one vertex per position; the right choice is the occurrence that is \emph{hardest to reach}: for each position we keep a vertex with the maximum threshold, and the maximality is what makes the cycles of the compressed graph positive.

\begin{proposition}[Compression]
\label{prop:compression}
A duplicate-free backward {\thrres} proof of size $S$ yields a refutation strategy graph for $F$ with at most $S$ vertices.
\end{proposition}
\begin{proof}
Call the coefficient vector of $f$ the \emph{position} of a vertex labeled $f \leq t$. For every position $p$ occurring in the proof, choose among the vertices of the proof at the position $p$ one with the maximum threshold; call it the \emph{representative} of $p$ and denote its threshold by $\rho(p)$. Since the backward proof is duplicate-free, the vertex with label $f_p\leq\rho(p)$ is unique. Let $Q$ be the graph with one vertex per occupied position, where the vertex at $p$ carries the move type of its representative and, for each proof edge from the representative to a child labeled $f_{p'} \leq s'$, an edge $p \to p'$ of weight $\rho(p) - s'$; root $Q$ at the origin (the position of the start vertex) and keep the part reachable from the root.

Every vertex of $Q$, with its edges, is of the type of its representative in Table~\ref{tab:local-types}: the moves are determined by the position and the type, and the edge weights $\rho(p) - s'$ are the threshold differences prescribed by the move. The terminal is never a representative, since the start vertex occupies the same position $\vec 0$ with the larger threshold $0 > -1$; hence every vertex of $Q$ inherits the moves of a non-terminal vertex and has out-degree at least $1$.

It remains to check that the directed cycles of $Q$ are strictly positive. Consider a cycle $p_1, p_2, \ldots, p_q, p_1$ (indices modulo $q$). The edge from $p_i$ to $p_{i+1}$ is the projection of a proof edge from the representative of $p_i$ to some vertex at $p_{i+1}$ with a threshold $s_{i+1} \leq \rho(p_{i+1})$, and its weight is $\rho(p_i) - s_{i+1}$. Summing around the cycle and shifting the index,
\[
\sum_{i=1}^{q} \big(\rho(p_i) - s_{i+1}\big) \;=\; \sum_{i=1}^{q} \big(\rho(p_i) - s_i\big) \;\geq\; 0 .
\]
If the total weight were $0$, every $s_i$ would equal $\rho(p_i)$: every edge of the cycle would end at the representative of the next position, and the same edges would form a directed cycle in the proof DAG, which is acyclic. Hence every cycle of $Q$ is strictly positive, and $Q$ is a refutation strategy graph for $F$ with at most $S$ vertices.
\end{proof}

\subsection{Assembly}
\label{sec:assembly}

\begin{proof}[Proof of the Main Lemma (Lemma~\ref{lm:main})]
The two directions compose the preceding statements; we only track the parameters.

\emph{From {\mpnsr} to {\thrres}.} An {\mpnsr} refutation of size $S$ yields, by Theorem~\ref{thm:refut-subgraph-equiv}, a refutation subgraph with $\poly(S, N)$ vertices, and, by the normal-form lemma (Lemma~\ref{lm:normal-form}), a refutation strategy graph $G$ with $r = \poly(S, N)$ vertices; its edge weights lie in $\{-1, 0, 1\}$, so $W' = 1$. Its positions have $\ell_1$-norm at most $P \leq r A = \poly(S, N)$, where $A = \poly(N)$ bounds the width of the clauses of $F$ and $A \geq 2$: every vertex of $G$ is reachable from the origin by a simple path of at most $r$ edges, and, by Table~\ref{tab:local-types}, every edge changes the position by at most $A$ in $\ell_1$-norm. The Unfolding Lemma (Lemma~\ref{lm:unfolding-lemma}) and the reversal (Proposition~\ref{prop:reversal}) then give a {\thrres} refutation of size $\poly(r, 1, P) = \poly(S, N)$; its coefficient vectors have $\ell_1$-norm at most $P + 2$ (positions of $G$, plus one complementary pair inside the cleanup chains) and its thresholds are $O(r)$ in absolute value, so the refutation is unary.

\emph{From {\thrres} to {\mpnsr}.} A {\thrres} refutation of size $S$, trimmed to the ancestors of the line $0 \geq 1$, becomes a backward proof of size at most $S$ (Proposition~\ref{prop:reversal}), then a refutation strategy graph with at most $S$ vertices (Proposition~\ref{prop:compression}), whose positions are coefficient vectors of the proof, of $\ell_1$-norm at most $S$; then a refutation subgraph of the monomial graph of polynomially related size (Lemma~\ref{lm:normal-form}), whose monomial labels have degree at most $S$ and whose edge weights lie in $\{-1, 0, 1\}$; and finally an {\mpnsr} refutation of size $\poly(S)$ (Theorem~\ref{thm:refut-subgraph-equiv}).
\end{proof}

\subsection{From daglike threshold proofs to treelike tropical proofs}
\label{sec:daglike-to-treelike}

A simple simulation complements the Main Lemma: the daglike system {\thrres}, whose lines may be reused arbitrarily, can be simulated by the \emph{treelike} dynamic tropical system. The simulation is not needed for the equivalence of {\mpnsr} and treelike {\mppcr}, which is known from~\cite{AGH25} (see Section~\ref{sec:triangle}); it is a property of the new system, recorded because its proof is short, because it extends uniformly to arbitrary classes of tropical axioms (Appendix~\ref{sec:appendix-generalized}), and because its pattern --- applying one premise of the refuted system inside an arbitrary context --- reappears in the simulation of Lift \& Project (Theorem~\ref{thm:mppcr-simulates-lp}). It rests on the observation that every rule of {\thrres} is such an application.

\begin{lemma}[Rules are premise applications]
\label{lm:rules-are-applications}
Let $F$ be a CNF formula over the variables $x_1, \ldots, x_N$. Every rule of {\thrres} for $F$ is an application, in an arbitrary context, of either a clause encoding~\eqref{eq:encoding-i} of $F$ or one of the standard and resolution axioms. Here, for a tropical premise
\[
\bigoplus_{r \in R} (a_r^T x - b_r) \;\leq\; c^T x - d,
\qquad a_r, c \in \Z_+^{2N},\ b_r, d \in \Z,
\]
its \emph{application in the context} $f - s$ derives
\[
f + c^T x \;\geq\; s + d
\]
from the lines
\[
f + a_r^T x \;\geq\; s + b_r, \qquad r \in R.
\]
The correspondence between the five rules of Definition~\ref{def:thrres} and the premises is as follows:
\begin{center}
\begin{tabular}{lll}
rule & applied premise & application in the context $f - s$\\
\hline
{\rulecl} & $1 \leq \bigodot_{l \in C} l$ & from $f \geq s - 1$ derive $f + \sum_{l \in C} l \geq s$\\
{\ruleres} & $x_i \oplus \overline{x_i} \leq 0$ & from $f + x_i \geq s$, $f + \overline{x_i} \geq s$ derive $f \geq s$\\
{\ruleweak} & $0 \leq x_i$ (or $0 \leq \overline{x_i}$) & from $f \geq s$ derive $f + l \geq s$\\
{\ruleintro} & $1 \leq x_i \odot \overline{x_i}$ & from $f \geq s - 1$ derive $f + x_i + \overline{x_i} \geq s$\\
{\rulesim} & $x_i \odot \overline{x_i} \leq 1$ & from $f + x_i + \overline{x_i} \geq s$ derive $f \geq s - 1$\\
\end{tabular}
\end{center}
\end{lemma}
\begin{proof}
Direct comparison of each row with the definition of an application. For {\rulecl}, the encoding $1 \leq \bigodot_{l \in C} l$ has the single left-hand term $0^T x - (-1)$ and the right-hand term $\big(\sum_{l \in C} e(l)\big)^T x - 0$, where $e(l)$ is the exponent vector of the literal $l$; its application derives $f + \sum_{l \in C} l \geq s + 0$ from $f \geq s + (-1)$, which is the {\rulecl} rule up to renaming the threshold. The other four rows are checked in the same way from Definition~\ref{def:mpnsr}.
\end{proof}

\begin{theorem}[Daglike to treelike]
\label{thm:daglike-to-treelike}
Unary treelike {\mppcr} polynomially simulates {\thrres}: a {\thrres} refutation of size $S$ can be transformed into a unary treelike {\mppcr} refutation of size $\poly(S)$.
\end{theorem}
\begin{proof}
List the nodes of the refutation as $\ell_1, \ldots, \ell_K$ in a topological order, with $\ell_1$ the source $0 \geq 0$; if $\ell_i = (a_i^T x \geq b_i)$, write $T_i := a_i^T x - b_i$, so that $0 \leq T_i$ is exactly $\ell_i$. We process the lines in this order and maintain the treelike {\mppcr} line
\[
V_i: \qquad 0 \;\leq\; T_1 \oplus T_2 \oplus \cdots \oplus T_i .
\]
Since tropical addition is the minimum, $V_i$ simultaneously records all lines processed so far: it is equivalent to the conditions $0 \leq T_j$ for every $j \leq i$. The initial line $V_1 = (0 \leq 0)$ is trivial.

Suppose the line $\ell_i$ is derived by a rule from the lines $\ell_{j_r}$, $r \in R$, with $j_r < i$. By Lemma~\ref{lm:rules-are-applications}, the rule is the application, in some context $f - s$, of a premise $\bigoplus_{r \in R} (a_r^T x - b_r) \leq c^T x - d$ of the refuted system; writing $h := f - s$ for the term of the context, the terms of the lines involved are exactly
\[
T_{j_r} = h \odot (a_r^T x - b_r) \quad (r \in R), \qquad T_i = h \odot (c^T x - d).
\]
Tropical multiplication of the premise by the term $h$ therefore yields precisely
\[
\bigoplus_{r \in R} T_{j_r} \;\leq\; T_i ,
\]
and combining it by the minimum rule with a fresh reflexive derivation $\bigoplus_{r \in R} T_{j_r} \leq \bigoplus_{r \in R} T_{j_r}$ (each $T_{j_r} \leq T_{j_r}$ is the trivial axiom $0 \leq 0$ tropically multiplied by $T_{j_r}$) gives
\[
\bigoplus_{r \in R} T_{j_r} \;\leq\; \bigoplus_{r \in R} T_{j_r} \oplus T_i .
\]
Substituting the displayed inequality into $V_{i-1}$ on the right (Remark~\ref{rem:substitutions}) replaces the subminimum $\bigoplus_{r\in R}T_{j_r}$ by $\bigoplus_{r\in R}T_{j_r}\oplus T_i$ and yields $V_i$.

When the contradiction $0 \geq 1$ is processed, its term $T_i$ is the constant $-1$; dropping all the other summands from the right-hand side of $V_i$ by natural weakening (\cite{AGH25} Note~2.14) yields the contradictory final line $0 \leq -1$.

The construction is treelike: the lines $V_1, V_2, \ldots$ form one main spine, each preceding spine line is used exactly once, every premise and reflexivity derivation attached to a stage is a fresh copy, and the premise terms $T_{j_r}$ enter a stage only as summands of the single line $V_{i-1}$, not as separate proof nodes --- so reusing a line of the daglike refutation many times does not reuse any line of the treelike derivation.

Finally, the size. There are $K \leq S$ stages, each costing $O(|R|)$ inferences with $|R|$ bounded by the width of the applied premise; the sums $V_i$ grow linearly with the number of processed lines, and the coefficients and exponent vectors are taken from the threshold lines or their rule contexts. Thus the total unary size of the resulting derivation is $\poly(S)$.
\end{proof}

\begin{example}[The toy refutation, translated]
\label{ex:daglike-to-mppcr}
Let $F = x_1 \wedge \overline{x_1}$, with the clauses encoded by $1 \leq x_1$ and $1 \leq \overline{x_1}$, and consider the four-line {\thrres} refutation of Example~\ref{ex:toy-thrres}. Its tropical terms are $T_1 = 0$, $T_2 = x_1 - 1$, $T_3 = \overline{x_1} - 1$, $T_4 = -1$, and the simulation maintains the treelike {\mppcr} lines
\begin{gather*}
V_1 = \big(0 \leq 0\big), \qquad V_2 = \big(0 \leq 0 \oplus (x_1 - 1)\big), \qquad V_3 = \big(0 \leq 0 \oplus (x_1 - 1) \oplus (\overline{x_1} - 1)\big),\\
V_4 = \big(0 \leq 0 \oplus (x_1 - 1) \oplus (\overline{x_1} - 1) \oplus (-1)\big).
\end{gather*}
The step to $V_2$ substitutes $0 \leq 0 \oplus (x_1 - 1)$, obtained from the clause encoding $1 \leq x_1$ tropically multiplied by $-1$ and combined with the reflexive $0 \leq 0$; the step to $V_3$ is symmetric. The step to $V_4$ applies the resolution axiom $x_1 \oplus \overline{x_1} \leq 0$ in the context $0 - 1$: tropically multiplied by the term $h = -1$, the axiom reads $(x_1 - 1) \oplus (\overline{x_1} - 1) \leq -1$, and combining it with the reflexive $(x_1 - 1) \oplus (\overline{x_1} - 1) \leq (x_1 - 1) \oplus (\overline{x_1} - 1)$ and substituting into $V_3$ yields $V_4$. Finally, dropping the summands $0$, $x_1 - 1$ and $\overline{x_1} - 1$ from the right-hand side of $V_4$ by natural weakening leaves the contradictory final line $0 \leq -1$.
\end{example}

\subsection{The equivalence triangle}
\label{sec:triangle}

The equivalence of {\mpnsr} and treelike {\mppcr} was proved in~\cite{AGH25} by direct simulations in both directions. The nontrivial direction is the flattening theorem below, which we use unchanged; the converse direct simulation --- a static combination is assembled term by term by a treelike derivation --- is elementary and also known. What we record here is a cycle of simulations that passes through the new system {\thrres}.

\begin{theorem}[Flattening treelike proofs, \cite{AGH25} Theorem~4.1]
\label{thm:treelike-to-ns}
Binary {\mpnsr} polynomially simulates binary treelike {\mppcr}.
\end{theorem}
\begin{proof}
Theorem~4.1 of~\cite{AGH25} states the simulation of treelike {\mppc} by {\mpns} for an arbitrary set of premises; apply it to the premise set of the {\mpnsr} encoding.
\end{proof}

We record the following cycle of simulations.

\begin{corollary}[Equivalence triangle, standard setting]
\label{cor:triangle-standard}
Binary {\mpnsr}, binary treelike {\mppcr} and unary {\thrres} polynomially simulate one another. The simulations form the cycle
\[
\text{treelike {\mppcr}}
\xrightarrow{\text{\cite[T4.1]{AGH25}}}
{\mpnsr}
\xrightarrow{\text{L\ref{lm:main}}}
\text{{\thrres}}
\xrightarrow{\text{T\ref{thm:daglike-to-treelike}}}
\text{treelike {\mppcr}} .
\]
\end{corollary}

None of the three simulations of the cycle depends on the specific choice of the resolution axioms: the same arguments apply over any unary class of tropical axioms, and the resulting general equivalence triangle is stated in Appendix~\ref{sec:appendix-generalized} (Theorem~\ref{thm:triangle}).

\section{Connections to Cutting Planes and Lift \& Project}
\label{sec:lp-connection}

In this section we relate the tropical systems to proof systems based on linear inequalities over the plain (not dual-encoded) Boolean variables. The connection goes in both directions. Reading a threshold condition as a linear inequality embeds {\thrres}, and hence all the systems of Corollary~\ref{cor:triangle-standard}, into unary daglike Cutting Planes and Lift \& Project, yielding lower bounds for the tropical systems. In the converse direction, binary daglike {\mppcr} simulates daglike {\LnP}, and simulates daglike {\CP} when division axioms are added.

We first define the two systems.

\begin{definition}[Cutting Planes~\cite{CCT87}]
\label{def:cutting-planes}
The {\CP} proof system works with linear inequalities $c^T x \geq d$, where $c \in \Z^N$ and $d \in \Z$, over the variables $x_1, \ldots, x_N$. The derivation rules are:
\begin{itemize}
\item \textbf{Linear combination}: from $c_1^T x \geq d_1$ and $c_2^T x \geq d_2$ derive $(\lambda_1 c_1 + \lambda_2 c_2)^T x \geq \lambda_1 d_1 + \lambda_2 d_2$, where $\lambda_1, \lambda_2 \in \Z_+$;
\item \textbf{Division (rounding)}: from $\alpha\, c^T x \geq d$, where $\alpha \in \Z$, $\alpha > 0$, derive $c^T x \geq \lceil d/\alpha \rceil$.
\end{itemize}
A Boolean clause is embedded as a linear inequality: for instance, $x_1 \vee x_2 \vee \neg x_3$ becomes $x_1 + x_2 + (1 - x_3) \geq 1$. The \emph{box axioms} $x_i \geq 0$ and $x_i \leq 1$ are added to enforce that the variables are Boolean. A refutation derives the contradiction $0 \geq 1$. Both daglike and treelike versions exist; some formulas are known to require exponentially long daglike {\CP} proofs~\cite{Pudlak97}.
\end{definition}

\begin{definition}[Lift \& Project~\cite{HK06}]
\label{def:lift-project}
We use the integer-coefficient presentation of the Lift \& Project proof system of Hirsch and Kojevnikov~\cite{HK06}, based on the lift-and-project method of Balas, Ceria and Cornu\'ejols~\cite{BCC93}. The {\LnP} proof system works with the same lines, clause embedding, box axioms and goal as {\CP}, and with the same linear combination rule; the division rule is replaced by a lifting rule for a single variable:
\begin{itemize}
\item \textbf{Lift \& Project rule}: from $c^T x + d' \geq d$ and $c^T x + d'' x_k \geq d$ derive $c^T x + d' x_k \geq d$, where $d', d'' \in \Z$. The rule is sound for $x_k \in \{0,1\}$: for $x_k = 0$ the conclusion is implied by the second premise ($d'' x_k = d' x_k = 0$), and for $x_k = 1$ by the first ($d' \geq d' x_k$ becomes an equality).
\end{itemize}
\end{definition}

In both systems, the size of a proof is the number of lines plus the total bitsize of the coefficients; in the \emph{unary} versions the coefficients are encoded in unary and their unary lengths are counted in the size, as in Section~\ref{sec:notation}. Note that the trivial line $0 \geq 0$ is available in both systems as the linear combination of any two premises with $\lambda_1 = \lambda_2 = 0$.

\subsection{Simulating {\thrres} in unary daglike CP and L\&P}
\label{sec:thrres-in-lp}

Threshold conditions are linear inequalities over the dual-encoded variables, so a threshold proof translates into an LP proof by eliminating the complementary variables: substituting $\overline{x_i} = 1 - x_i$ turns a line $f \geq s$ into an inequality $c^T x \geq d$ over the plain variables, with $\|c\|_\infty$ bounded by the multiplicities of $f$ and $|d| \leq |s| + |f|$. We denote this translation by $f \geq s \mapsto \langle f \geq s \rangle$; the translation of the encoding~\eqref{eq:encoding-i} of a clause is exactly its {\CP} embedding.

\begin{lemma}[Unary daglike CP and L\&P simulate {\thrres}]
\label{lm:lp-cp-simulate-thrres}
Unary daglike {\CP} and unary daglike {\LnP} each polynomially simulate {\thrres}.
\end{lemma}
\begin{proof}
Translate every line of the refutation by $\langle\cdot\rangle$ and process the rules in a topological order; the required steps are listed in Table~\ref{tab:thrres-in-lp} and verified below.

\begin{table}[ht]
\centering
\footnotesize
\begin{tabular}{lll}
{\thrres} step & in {\CP} & in {\LnP}\\
\hline
source $0 \geq 0$ & trivial combination ($\lambda_1 = \lambda_2 = 0$) & same as {\CP}\\
{\rulecl} & add the clause embedding & same as {\CP}\\
{\ruleweak} & add a box axiom & same as {\CP}\\
{\ruleintro}, {\rulesim} & nothing (the translation is unchanged) & same as {\CP}\\
{\ruleres} & add the premises, divide by $2$ & the Lift \& Project rule\\
\end{tabular}
\caption{Simulating the rules of {\thrres} in unary daglike {\CP} and {\LnP}.}
\label{tab:thrres-in-lp}
\end{table}

Since $[x_i] + [\overline{x_i}] = 1$ identically after the substitution, the rules {\ruleintro} and {\rulesim} do not change the translated line and are skipped. A {\rulecl} step is the linear combination of the current line with the embedding $\langle 1 \leq \bigodot_{l \in C} l\rangle = \big(\sum_{l \in C} [l] \geq 1\big)$ of the input clause, and a {\ruleweak} step is the linear combination with the box axiom $[l] \geq 0$ (which is $x_i \geq 0$ or $1 - x_i \geq 0$).

For {\ruleres}, let the translated premises be $a^T x + x_k \geq s$ and $a^T x + (1 - x_k) \geq s$ (the translations of $f + x_k \geq s$ and $f + \overline{x_k} \geq s$). In {\CP}, their sum is $2a^T x + 1 \geq 2s$, i.e.\ $2a^T x \geq 2s - 1$, and division by $2$ yields $a^T x \geq \lceil s - 1/2\rceil = s$, the translated conclusion. In {\LnP}, write $a' := a - e_k$: the first premise reads $a'^T x + 2 x_k \geq s$ and the second $a'^T x + 1 \geq s$, so the Lift \& Project rule with $d' = 1$, $d'' = 2$ derives $a'^T x + x_k \geq s$, which is again the translated conclusion.

All coefficients appearing in the simulation are bounded by the multiplicities and thresholds of the original proof, hence remain polynomially bounded: the simulations are unary.
\end{proof}

\begin{corollary}
\label{cor:mpnsr-in-lp}
Unary daglike {\CP} and unary daglike {\LnP} polynomially simulate binary {\mpnsr} and binary treelike {\mppcr}.
\end{corollary}
\begin{proof}
Combine Lemma~\ref{lm:lp-cp-simulate-thrres} with Corollary~\ref{cor:triangle-standard}.
\end{proof}

\begin{corollary}[Lower bounds for the tropical systems]
\label{cor:tropical-lower-bounds}
The standard clique-colouring CNFs asserting that a graph on $r$ vertices contains a clique of size $q=\lfloor\tfrac18(r/\log r)^{2/3}\rfloor$ and has a $(q-1)$-colouring require refutations of size $2^{\Omega((r/\log r)^{1/3})}$ in unary {\thrres}, binary {\mpnsr} and binary treelike {\mppcr}.
\end{corollary}
\begin{proof}
The corresponding {\CP} lower bound is due to Pudl\'ak~\cite[Corollary~7]{Pudlak97}. It transfers through Lemma~\ref{lm:lp-cp-simulate-thrres} and Corollary~\ref{cor:mpnsr-in-lp}; the polynomial simulation overhead preserves the stated bound.
\end{proof}

In the same way, any lower bound on the size of unary daglike {\CP} or {\LnP} refutations of CNFs transfers to these three systems.

\subsection{Binary daglike simulations}
\label{sec:binary-simulations}

We now turn to the converse direction: simulating daglike {\CP} and {\LnP} in tropical systems. Coefficients in daglike {\CP} and {\LnP} proofs may be binary, so any simulation must be able to \emph{produce} binary coefficients. In the daglike tropical setting this is provided by tropical products and powers.

\begin{remark}[Tropical products and powers, daglike]
\label{rem:products}
From $p \leq q$ and $p' \leq q'$ one can derive $p \odot p' \leq q \odot q'$ (\cite[Note~2.13]{AGH25}): multiply the first premise by every monomial of $p'$ and combine the results by the minimum rule, obtaining $p \odot p' \leq q \odot p'$; likewise multiply the second premise by every monomial of $q$, obtaining $q \odot p' \leq q \odot q'$; conclude by transitivity. The derivation uses the first premise once per monomial of $p'$ and the second once per monomial of $q$, so it is efficient in the \emph{daglike} setting. For tropical terms $A,B$, iterated squaring then yields tropical powers: from $A \leq B$ one derives $A^{\odot \lambda} \leq B^{\odot \lambda}$ in $O(\log^2 \lambda)$ inferences (\cite[Lemma~3.6]{AGH25}). Since the tropical power of a linear inequality $0 \leq g$ is its scalar multiple $0 \leq \lambda g$, this provides multiplication of derived lines by binary constants at polynomial cost.
\end{remark}

For nonnegative $d', d''$, the Lift \& Project rule corresponds to the tropical inequality
\[
d' \oplus x_k^{\odot d''} \;\leq\; x_k^{\odot d'} , \qquad d', d'' \in \Z_{\geq 0}
\]
(the \emph{Lift \& Project inequality}; it is valid over the Boolean points: for $x_k = 0$ both sides evaluate to $0$, and for $x_k = 1$ it reads $\min\{d', d''\} \leq d'$). Unlike the division axioms below, it need not be added as an axiom: it is derivable.

\begin{lemma}[The Lift \& Project inequality is derivable]
\label{lm:lp-axiom-from-res}
For $d', d'' \in \Z_{\geq 0}$, binary daglike {\mppcr} derives $d' \oplus x_k^{\odot d''} \leq x_k^{\odot d'}$ from the standard and resolution axioms at cost polynomial in the bitsizes of $d'$ and $d''$.
\end{lemma}
\begin{proof}
Tropical powers of the axioms $0 \leq x_k$ and $1 \leq x_k \odot \overline{x_k}$ (Remark~\ref{rem:products}) give, for any $a \geq 1$,
\[
0 \;\leq\; x_k^{\odot a}
\qquad\text{and}\qquad
a \;\leq\; x_k^{\odot a} \odot \overline{x_k}^{\odot a}
\]
in $O(\log^2 a)$ inferences. We also need the mixed powers $E_{a,b} := \big(x_k^{\odot a} \oplus \overline{x_k}^{\odot b} \leq 0\big)$ of the resolution axiom $E_{1,1}$; the exponent $a$ is doubled in four inferences (and $b$ symmetrically):
\begin{gather*}
x_k^{\odot 2a} \oplus \overline{x_k}^{\odot b} \odot x_k^{\odot a} \leq x_k^{\odot a}
\quad\text{($E_{a,b}$ times $x_k^{\odot a}$)},\\
\overline{x_k}^{\odot b} \leq \overline{x_k}^{\odot b} \odot x_k^{\odot a}
\quad\text{($0 \leq x_k^{\odot a}$ times $\overline{x_k}^{\odot b}$)},
\end{gather*}
so substitution gives $x_k^{\odot 2a} \oplus \overline{x_k}^{\odot b} \leq x_k^{\odot a}$, and the minimum rule with $\overline{x_k}^{\odot b} \leq \overline{x_k}^{\odot b}$ followed by transitivity with $E_{a,b}$ gives $E_{2a,b}$. Each doubling uses $E_{a,b}$ twice, which is allowed in a daglike proof, and double-and-add chains of such steps produce $E_{a,b}$ for arbitrary $a, b \geq 1$ in $O(\log a + \log b)$ inferences.

Now derive $d' \oplus x_k^{\odot d''} \leq x_k^{\odot d'}$. For $d'=0$, the claim follows by the minimum rule from $0\leq0$ and $x_k^{\odot d''}\leq\infty$; hence assume $d'>0$. If $d'' \leq d'$: multiply $0 \leq x_k^{\odot(d' - d'')}$ by $x_k^{\odot d''}$ to get $x_k^{\odot d''} \leq x_k^{\odot d'}$, and add the constant term on the left by the minimum rule with the trivial axiom $d' \leq \infty$. If $d'' > d'$: multiply $E_{d''-d',\,d'}$ by $x_k^{\odot d'}$ to get $x_k^{\odot d''} \oplus \overline{x_k}^{\odot d'} \odot x_k^{\odot d'} \leq x_k^{\odot d'}$, and substitute $d' \leq x_k^{\odot d'} \odot \overline{x_k}^{\odot d'}$ into the second term on the left. In both cases the total cost is polynomial in the bitsizes of $d'$ and $d''$.
\end{proof}

The Lift \& Project rule of Definition~\ref{def:lift-project} permits arbitrary integers $d', d''$, while the Lift \& Project inequality is available for nonnegative coefficients only. The gap is closed inside {\LnP} itself: an application with $d' \geq d''$ or with $d' < 0$ is not a genuine lift.

\begin{lemma}[Normalization of lift applications]
\label{lm:lp-normalization}
In a daglike {\LnP} proof, every application of the Lift \& Project rule with $d' \geq d''$ or with $d' < 0$ can be replaced by a single linear combination with a box axiom. Consequently, every {\LnP} refutation can be transformed, with at most a constant-factor increase in the number of lines and an additive $O(1)$ increase in the maximum coefficient bitsize (including rule multipliers), into one in which every application of the rule has $0 \leq d' < d''$.
\end{lemma}
\begin{proof}
If $d' \geq d''$, the conclusion $c^T x + d' x_k \geq d$ is the linear combination of the second premise $c^T x + d'' x_k \geq d$ with the box axiom $x_k \geq 0$ taken with the multiplier $\lambda = d' - d'' \geq 0$. If $d' < 0$, the conclusion is the linear combination of the first premise $c^T x + d' \geq d$ with the box axiom $1 - x_k \geq 0$ taken with the multiplier $\lambda = -d' > 0$: indeed, $(c^T x + d') + (-d')(1 - x_k) = c^T x + d' x_k$. Every remaining application satisfies $d' \geq 0$ and $d' < d''$. The new multiplier $d'-d''$ or $-d'$ has bitsize at most a constant plus the larger of the original bitsizes; any needed box-axiom line adds only constant overhead per step.
\end{proof}

\begin{theorem}[Binary daglike {\mppcr} simulates daglike L\&P]
\label{thm:mppcr-simulates-lp}
Binary daglike {\mppcr} (with the standard axioms, the resolution axioms and the clause encodings~\eqref{eq:encoding-i}) polynomially simulates daglike {\LnP}.
\end{theorem}
\begin{proof}
The invariant of the simulation: every {\LnP} line $u^T x \geq v$ is maintained as the tropical line $0 \leq u^T x - v$ in the dual encoding, where a negative coefficient at a variable is first rewritten through $x_i = 1 - \overline{x_i}$, so that it becomes a nonnegative coefficient at the complementary variable with the constant term adjusted. The resulting term is normalized: no variable occurs in both polarities. Powers of the pair axioms, $m\leq(x_i\odot\overline{x_i})^{\odot m}$ and $(x_i\odot\overline{x_i})^{\odot m}\leq m$, allow us to insert or cancel complementary pairs by multiplication and transitivity, at polynomial cost in their binary encoding (Remark~\ref{rem:products}). Input clauses are their encodings~\eqref{eq:encoding-i}, and the box axioms $x_i \geq 0$, $x_i \leq 1$ are the standard axioms $0 \leq x_i$, $0 \leq \overline{x_i}$ in the dual encoding.

A \emph{linear combination} with multipliers $\lambda_1, \lambda_2$ is simulated by the tropical powers $0 \leq \lambda_j(u_j^T x - v_j)$ of the maintained lines followed by their tropical product (Remark~\ref{rem:products}), in $O(\log^2 \lambda_1 + \log^2 \lambda_2)$ inferences; a zero multiplier is replaced by the trivial axiom $0 \leq 0$, and the resulting term is normalized by cancelling complementary pairs.

By Lemma~\ref{lm:lp-normalization} we may assume that every application of the \emph{Lift \& Project rule}, from $a^T x + d' \geq b$ and $a^T x + d'' x_k \geq b$ derive $a^T x + d' x_k \geq b$, has $0 \leq d' < d''$. Let $h$ be the normalized dual-encoded term representing $a^T x-b$. Using the pair-axiom rewritings above, put the maintained premises into the forms $0\leq h\odot d'$ and $0\leq h\odot x_k^{\odot d''}$ with this common $h$. Multiplying the Lift \& Project inequality of Lemma~\ref{lm:lp-axiom-from-res} by $h$ gives
\[
(h \odot d') \oplus \big(h \odot x_k^{\odot d''}\big) \;\leq\; h \odot x_k^{\odot d'} .
\]
The maintained premises are $0 \leq h \odot d'$ and $0 \leq h \odot x_k^{\odot d''}$; the minimum rule combines them into $0 \leq (h \odot d') \oplus \big(h \odot x_k^{\odot d''}\big)$, and transitivity with the displayed inequality yields $0 \leq h \odot x_k^{\odot d'}$, which is normalized by cancelling complementary pairs to obtain the maintained form of the conclusion.

The final {\LnP} line $0 \geq 1$ is maintained as $0 \leq -1$, a contradictory line. The powers and pair-axiom rewritings have polynomial cost in the binary size of the simulated step, and all coefficients and exponents have polynomial bitsize.
\end{proof}

For {\CP} the lifting rule is replaced by division, which the tropical rules do not provide; we add it as an axiom scheme.

\begin{definition}[The division axioms]
\label{def:division-axioms}
For $a \in \Z_+^{2N}$, $d \in \Z$ and $\alpha \in \Z_{\geq 1}$, the \emph{division axiom} is the tropical inequality
\[
\min\big\{\, 0,\ \alpha\, a^T x - \alpha d - 1 \,\big\} \;\leq\; a^T x - d - 1
\]
over the dual-encoded variables. Division axioms are valid at every Boolean point: there $a^T x$ is a nonnegative integer, and if $a^T x \geq d + 1$ the right-hand side is nonnegative, while if $a^T x \leq d$ then, writing $u = a^T x - d \leq 0$, the second term on the left equals $\alpha u - 1 \leq u - 1$, which is smaller than the right-hand side $u - 1$ or equal to it. Adding valid inequalities as premises preserves soundness of the refutations, and membership in the scheme is decidable in polynomial time.
\end{definition}

\begin{theorem}[Binary daglike {\mppcr} with division simulates daglike CP]
\label{thm:mppcr-simulates-cp}
Binary daglike {\mppcr} with the standard axioms, the resolution axioms, the clause encodings~\eqref{eq:encoding-i} and the division axioms polynomially simulates daglike {\CP}.
\end{theorem}
\begin{proof}
Maintain {\CP} lines and simulate linear combinations exactly as in the proof of Theorem~\ref{thm:mppcr-simulates-lp}, including the pair-axiom rewritings. Normalizing $\alpha u^T x-v$ shifts $v$ by a multiple of $\alpha$, so it commutes with division and rounding. For a \emph{division} step, write the maintained premise in normalized dual form as $0 \leq \alpha u^T x - v$, with $u\in\Z_+^{2N}$, and let $d + 1 = \lceil v / \alpha \rceil$, so that $\alpha d + 1 \leq v$. First weaken the premise to $0 \leq \alpha u^T x - \alpha d - 1$: the constant $c := v - \alpha d - 1$ is nonnegative, $0 \leq c$ is a tropical combination of the trivial axioms (the minimum of $0 \leq \infty$ and $0 \leq 0$ multiplied by $c$), and multiplying $0 \leq c$ by the term $\alpha u^T x - v$ followed by transitivity gives the weakened line. Now apply the division axiom with $a := u$: the minimum rule combines $0 \leq 0$ and the weakened line into $0 \leq \min\{0, \alpha u^T x - \alpha d - 1\}$, and transitivity with the axiom yields $0 \leq u^T x - d - 1$, the maintained form of the conclusion $u^T x \geq d + 1 = \lceil v/\alpha\rceil$.

The final line and the size bookkeeping are as in Theorem~\ref{thm:mppcr-simulates-lp}: every step, including the pair-axiom rewritings, has polynomial cost in its binary size, and all coefficients and exponents have polynomial bitsize.
\end{proof}

\section{Bounds in the degree-of-falsity hierarchy}
\label{sec:hierarchy}

We study how restricting the thresholds affects the strength of {\thrres}. We first show that {\thrres} restricted to proofs of degree of falsity at most $1$ is polynomially equivalent to daglike Resolution (Section~\ref{sec:thrres-vs-resolution}). Combining the {\CP} simulation of Section~\ref{sec:thrres-in-lp} with a theorem of Hirsch and Nikolenko~\cite{DBLP:conf/sat/HirschN05}, we obtain a quantitative simulation by Resolution and, in particular, extend this equivalence to every constant level of the hierarchy (Theorem~\ref{thm:constant-k-collapse}). Finally, we give bounds for the pigeonhole principle and Tseitin formulas (Sections~\ref{sec:php-hierarchy} and~\ref{sec:tseitin-hierarchy}).

\subsection{Degree of falsity one and daglike Resolution}
\label{sec:thrres-vs-resolution}

\begin{theorem}
\label{thm:1-thrres-resolution}
{\thrres} restricted to proofs of degree of falsity at most $1$ is polynomially equivalent to daglike Resolution.
\end{theorem}
\begin{proof}
First, {\thrres} simulates daglike Resolution using proofs of degree of falsity at most $1$. A line $f \geq 1$ is the disjunction of the literals of $f$. A clause $C$ of the CNF is derived from the source $0 \geq 0$ by one {\rulecl} step, and ordinary weakening is {\ruleweak}. For a resolution step deriving $A \vee B$ from $A \vee x_i$ and $B \vee \overline{x_i}$ (where $f_D:=\sum_{l\in D}l$ for a clause $D$ viewed as a set of literals), apply {\ruleweak} only for the literals in $B\setminus A$ and $A\setminus B$, respectively, obtaining $f_{A\cup B}+x_i\geq1$ and $f_{A\cup B}+\overline{x_i}\geq1$, and apply {\ruleres}. The empty clause becomes the line $0 \geq 1$, and all thresholds are $0$ or $1$.

Conversely, consider a {\thrres} refutation of degree of falsity at most $1$. A line with a nonpositive threshold is trivially true, and so is a line $f \geq 1$ in which $f$ contains a complementary pair. Delete all trivially true lines, and read every remaining line $f \geq 1$ as the clause $C_f$ on the distinct literals of $f$. We check that the surviving lines form a daglike Resolution derivation with weakening:
\begin{itemize}
\item the conclusions of {\rulesim} have nonpositive threshold, and the conclusions of {\ruleintro} contain a complementary pair, so both were deleted;
\item a surviving {\rulecl} conclusion $f + \sum_{l \in C} l \geq 1$ (its premise has threshold $0$) is a weakening of the input clause $C$;
\item a surviving {\ruleweak} conclusion has a surviving premise (a trivially true premise would make the conclusion trivially true as well) and is a weakening of it;
\item for a surviving {\ruleres} conclusion $f \geq 1$ with premises $f + x_i \geq 1$ and $f + \overline{x_i} \geq 1$: if both premises survive, this is a resolution step on $C_{f + x_i}$ and $C_{f + \overline{x_i}}$; if a premise, say $f + x_i \geq 1$, was deleted, then $\overline{x_i}$ occurs in $f$, hence $C_{f + \overline{x_i}} = C_f$ and the conclusion coincides with the other premise as a clause.
\end{itemize}
The line $0 \geq 1$ becomes the empty clause. This yields a daglike Resolution refutation using the weakening rule, and Resolution with weakening is polynomially equivalent to plain Resolution.
\end{proof}

\begin{corollary}[Completeness]
\label{cor:thrres-complete}
For every $k\geq1$, {\thrres} restricted to proofs of degree of falsity at most $k$ is complete.
\end{corollary}
\begin{proof}
By Theorem~\ref{thm:1-thrres-resolution}, every Resolution refutation can be transformed into a {\thrres} refutation of degree of falsity at most $1$, and Resolution is complete. Such a refutation also has degree of falsity at most $k$ for every $k\geq1$.
\end{proof}

\subsection{Quantitative simulation by Resolution}
\label{sec:quantitative-resolution}

The {\CP} simulation of Lemma~\ref{lm:lp-cp-simulate-thrres} interacts well with the \emph{degree of falsity} of Goerdt~\cite{Goerdt91}: the constant term of a {\CP} inequality written with nonnegative coefficients over the literals (\cite[Lemma~1]{DBLP:conf/sat/HirschN05}; cf.\ Remark~\ref{rem:hkkn}). Hirsch and Nikolenko proved that a {\CP} proof $\Pi$ of a CNF in $N$ variables all whose lines have degree of falsity at most $k \leq N/2$ can be transformed into a resolution proof of size $O\big(\tbinom{N}{k-1}\,|\Pi|\,2^{6k}\big)$~\cite[Theorem~1]{DBLP:conf/sat/HirschN05}; the rules of Definition~\ref{def:cutting-planes} are instances of the {\CP} rules used there. This collapses every constant level of the threshold hierarchy.

\begin{theorem}[Quantitative simulation by Resolution]
\label{thm:constant-k-collapse}
A {\thrres} refutation of a CNF on $N$ variables, of size $S$ and degree of falsity at most $k$, where $1\leq k\leq N/4$, can be transformed into a daglike Resolution refutation of size
\[
O\!\left(\binom{N}{2k-1}2^{12k}\,\poly(S)\right),
\]
where the polynomial in $S$ is independent of $k$ and $N$. In particular, for every fixed integer $k\geq1$, {\thrres} restricted to proofs of degree of falsity at most $k$ is polynomially equivalent to daglike Resolution.
\end{theorem}
\begin{proof}
Daglike Resolution is simulated by {\thrres} with degree of falsity at most $1$ (Theorem~\ref{thm:1-thrres-resolution}), hence at most $k$. For the converse, apply Lemma~\ref{lm:lp-cp-simulate-thrres} to a {\thrres} refutation of size $S$ and degree of falsity at most $k$, and inspect the degrees of falsity in the resulting {\CP} proof of size $\poly(S)$. A translated line $\langle f \geq s \rangle$ has degree of falsity at most $s \leq k$: in the normalized literal form the threshold only decreases, by $1$ per complementary pair of $f$. The clause embeddings and the box axioms have degree of falsity at most $1$. The only remaining lines are the sums of two translated premises inside the {\ruleres} steps, and the degree of falsity of a sum of two inequalities is at most the sum of their degrees of falsity, so these intermediate lines have degree of falsity at most $2k$. For $N\geq4k$, the theorem of Hirsch and Nikolenko with degree bound $2k$ gives the stated size bound; the polynomial overhead comes from Lemma~\ref{lm:lp-cp-simulate-thrres} and is uniform in $k$ and $N$. For fixed $k$, this bound is polynomial in $S$ and $N$; if $N<4k$, the number of variables is bounded by a constant, and exhaustive search gives a constant-size Resolution refutation.
\end{proof}

\subsection{The pigeonhole principle}
\label{sec:php-hierarchy}

\begin{proposition}[The pigeonhole principle in the hierarchy]
\label{prop:php-hierarchy}
For $n\geq2$, let $\mathrm{PHP}^{n+1}_n$ be the ordinary pigeonhole principle, with variables $p_{ij}$ for $i\in[n+1]$, $j\in[n]$, pigeon clauses $\bigvee_{j\in[n]}p_{ij}$ and hole clauses $\overline{p_{ij}}\vee\overline{p_{i'j}}$ for $i\ne i'$. Thus the number of variables is $N=n(n+1)$.
\begin{enumerate}
\item $\mathrm{PHP}^{n+1}_n$ has {\thrres} refutations of size $\poly(n)$ and degree of falsity at most $2n+1$.
\item There is a constant $\varepsilon>0$ such that every {\thrres} refutation of $\mathrm{PHP}^{n+1}_n$ of degree of falsity at most $\varepsilon n/\log n$ has size $2^{\Omega(n)}$.
\end{enumerate}
\end{proposition}
The upper-bound construction is the {\thrres} counterpart of the short-to-long-inequality argument of Alekseev, Grigoriev and Hirsch~\cite[Lemma~4.4]{AGH25}; here we also track the thresholds.

\begin{lemma}
\label{lm:php-hole}
Fix a hole $j$ and $1\leq k\leq n+1$. From any line $f\geq s$, {\thrres} derives
\begin{equation}
\label{eq:php-hole-induction}
f+\sum_{i=1}^k\overline{p_{ij}}\geq s+k-1
\end{equation}
in $O(k^2)$ additional lines, all with thresholds at most $s+k-1$.
\end{lemma}
\begin{proof}
For $k=1$, use {\ruleweak}. Assume the claim for $k$ and put $c_m=f+\sum_{i=1}^k\overline{p_{ij}}+m\overline{p_{k+1,j}}$. Apply the induction hypothesis~\eqref{eq:php-hole-induction} and then {\ruleintro} on $\overline{p_{k+1,j}},p_{k+1,j}$ to obtain
\begin{equation}
\label{eq:php-resolution-premise}
c_1+p_{k+1,j}\geq s+k.
\end{equation}
Separately, starting from $f\geq s$, apply {\rulecl} successively with the $k$ hole clauses $\overline{p_{ij}}\vee\overline{p_{k+1,j}}$ to obtain $c_k\geq s+k$. For $m=k-1,\ldots,1$, resolve $c_{m+1}=c_m+\overline{p_{k+1,j}}\geq s+k$ against $c_m+p_{k+1,j}\geq s+k$, obtaining $c_m\geq s+k$. The latter premise for each $m$ is obtained from~\eqref{eq:php-resolution-premise} by adding $m-1$ copies of $\overline{p_{k+1,j}}$ using {\ruleweak}; all these premises are built incrementally in $O(k)$ lines before the resolutions. The result is $c_1\geq s+k$, as required for $k+1$. All new thresholds are at most $s+k$, and the induction step adds $O(k)$ lines, giving $O(k^2)$ lines in total.
\end{proof}

\begin{proof}[Proof of Proposition~\ref{prop:php-hierarchy}]
For the upper bound, start from $0\geq0$ and apply {\rulecl} with all $n+1$ pigeon clauses, obtaining $\sum_{i,j}p_{ij}\geq n+1$. Process the holes $j=1,\ldots,n$ in order. Before hole $j$, the current line is
\[
\sum_{i=1}^{n+1}\sum_{h=j}^n p_{ih}\geq n+2-j.
\]
Apply Lemma~\ref{lm:php-hole} with $k=n+1$, adding $\sum_i\overline{p_{ij}}$ and increasing the threshold by $n$. Then apply {\rulesim} to the $n+1$ complementary pairs at hole $j$, removing them and decreasing the threshold by $n+1$. After hole $n$ the line is $0\geq1$. The largest threshold is $2n+1$, reached while processing the first hole. There are $O(n^3)$ lines, each of polynomial unary size.

For the lower bound, Haken's theorem~\cite{Hak85} gives a constant $a>0$ such that every Resolution refutation of $\mathrm{PHP}^{n+1}_n$ has size at least $2^{an}$ for sufficiently large $n$. A {\thrres} refutation of size $S$ and degree of falsity at most $k\leq\varepsilon n/\log n$ yields, by Theorem~\ref{thm:constant-k-collapse}, a Resolution refutation of size at most
\[
O\!\left(N^{2k}2^{12k}\,\poly(S)\right),\qquad N=n(n+1).
\]
For a sufficiently small constant $\varepsilon>0$, the factor $N^{2k}2^{12k}$ is at most $2^{an/2}$, and $k\leq N/4$. Consequently $\poly(S)\geq2^{\Omega(n)}$, whence $S\geq2^{\Omega(n)}$.
\end{proof}

\subsection{Tseitin formulas at linear degree of falsity}
\label{sec:tseitin-hierarchy}

\begin{proposition}[Tseitin formulas at linear degree of falsity]
\label{prop:tseitin-degree}
Let $G$ range over a fixed family of $d$-regular edge expanders, for a constant $d\geq3$. Let $T_G$ be the standard Tseitin CNF with an odd total charge: its $N=|E(G)|$ variables correspond to edges, and at each vertex it encodes that the parity of the incident edges equals the vertex charge. There is a constant $\delta>0$ such that every {\thrres} refutation of $T_G$ of degree of falsity at most $\delta N$ has size $2^{\Omega(N)}$.
\end{proposition}
\begin{proof}
The Resolution lower bound for Tseitin formulas on bounded-degree expanders~\cite{Urq87,BSW01} gives a constant $a>0$ such that every Resolution refutation of $T_G$ has size at least $2^{aN}$ for sufficiently large $N$. Write $H_2(t)=-t\log_2t-(1-t)\log_2(1-t)$ for binary entropy. Choose $0<\delta<1/4$ so small that $H_2(2\delta)+12\delta<a/2$.

If a {\thrres} refutation has size $S$ and degree of falsity $k\leq\delta N$, Theorem~\ref{thm:constant-k-collapse} yields a Resolution refutation of size
\[
O\!\left(\binom{N}{2k-1}2^{12k}\,\poly(S)\right)
\leq 2^{(H_2(2\delta)+12\delta)N}\,O(\poly(S)).
\]
Here we used $\binom{N}{j}\leq2^{NH_2(j/N)}$ and the monotonicity of $H_2$ on $[0,1/2]$. Comparing with $2^{aN}$ gives $\poly(S)\geq2^{\Omega(N)}$, hence $S\geq2^{\Omega(N)}$.
\end{proof}

\section{Open questions}
\label{sec:open}

The simulations in this paper suggest studying tropical systems as intermediate systems between static certificates and reasoning with disjunctions of linear inequalities. Three sources of strength deserve to be distinguished: compact scaling by binary constants, reuse of derived lines, and additional inference rules or splitting axioms.

\paragraph{Lower bounds for daglike tropical proofs.} Our lower bounds apply to {\mpnsr} and treelike {\mppcr}; a central next goal is to prove superpolynomial lower bounds for binary daglike {\mppcr} refutations of CNFs. Can the monomial-graph viewpoint yield a direct combinatorial method, rather than a transfer of lower bounds from {\CP}? Restricted-depth proofs or proofs with limited reuse of derived lines may provide intermediate targets. The longer-term goal is to develop methods that survive the addition of stronger rules: tropical extensions by {\trres}~\cite{AGH25} and splitting axioms provide a framework for approaching lower bounds for {\reslp} and {\rescp}.

Even within the static setting, quantitative questions remain. For $\mathrm{PHP}^{n+1}_n$, Proposition~\ref{prop:php-hierarchy} gives exponential lower bounds up to degree of falsity $\varepsilon n/\log n$ and polynomial-size proofs at degree $2n+1$. Where does the transition occur, and can analogous tradeoffs help analyse stronger tropical systems?

\paragraph{Scaling versus reuse of derived lines.} How much stronger is binary daglike {\mppcr} than {\mpnsr}, or equivalently treelike {\mppcr}, on CNFs? Outside CNFs the gap can be exponential: the tropical binomial $x_1^{\odot 2^n}=-1$ has short binary daglike {\mppcr} refutations, while its {\mpnsr} refutations have size greater than $2^n$~\cite[Section~7.5]{AGH25}. Is there a polynomial-time recognizable unary class of tropical axioms whose addition to binary {\mpnsr} allows it to polynomially simulate unary daglike {\LnP} or {\CP}? By Appendix~\ref{sec:appendix-generalized}, this question can equivalently be asked for generalized treelike {\mppcr} or the generalized daglike threshold system over the same axiom class.

A natural intermediate system retains static or treelike proofs but admits binary tropical powers of the input inequalities and Boolean axioms as additional premises. Such an extension isolates compact scaling from general daglike reuse and is motivated by the simulations of {\LnP} and, with division axioms, {\CP} in Section~\ref{sec:binary-simulations}. Its precise formulation and comparison with these linear systems deserve a separate treatment. In particular, can arbitrary daglike reuse be eliminated once such scaled premises are available, or can the resulting systems be separated? The unary-axiom-class hypothesis of Appendix~\ref{sec:appendix-generalized} does not cover these binary-scaled premises.

A separate question, related to the comparison with treelike {\CP} raised in~\cite[Section~8]{AGH25}, is whether {\mpnsr} polynomially, or at least quasipolynomially, simulates unary treelike {\CP} or unary treelike {\LnP}.

A related question is whether unary and binary daglike {\mppcr} are polynomially equivalent on CNFs. For the static and treelike systems they are (Corollaries~\ref{cor:unary-coefficients} and~\ref{cor:triangle-standard}), but our daglike simulations use binary powers essentially. Understanding this distinction would separate the role of numerical encoding from the role of proof structure.

\paragraph{Additional rules and splitting axioms.} What is the cost of eliminating {\trres} from daglike and from treelike tropical refutations of CNFs? The known obstruction to expressing the rule as a static tropical combination~\cite{AGH25} does not by itself give a separation between refutation sizes. In the daglike setting, adding this rule gives the known connection with {\reslp}; in the treelike setting, identifying the appropriate tropical counterparts of treelike {\reslp} and Stabbing Planes remains an important direction. A concrete intermediate question raised in~\cite[Section~8]{AGH25} is whether treelike {\reslp} polynomially simulates {\mpnsr}.
Can suitable tropical extensions clarify whether treelike {\reslp} (with binary coefficients) admits a quasipolynomial simulation in {\CP}?

It is also useful to distinguish Boolean splitting, real splitting $f\geq0\lor f<0$, and integer splitting $f\geq1\lor f\leq0$ for integer affine forms $f$. The equivalence of unary daglike {\reslp} and {\rescp}~\cite{HK06} does not automatically extend to binary or treelike proofs. Studying these choices in tropical systems may clarify the strength of disjunctive reasoning relative to {\CP} and {\LnP}. For {\rescp} versus {\CP}, the quantitative goal is to strengthen the superpolynomial separation known for suitably parameterized clique-colouring formulas~\cite{ABE02,Stabbing2} to separations with lower bounds $2^{N^{\Omega(1)}}$, or even $2^{\Omega(N)}$, against polynomial-size proofs, where $N$ is the number of variables.

\section*{Acknowledgements}
We thank Dima Grigoriev and Anat Paskin-Cherniavsky for their careful reading of the thesis of the second author (Dmitrii) on which this paper is based and for their helpful comments. Dmitrii is also deeply grateful to his mother, who for two years patiently listened to his barely intelligible rambling, diluted with mathematical terms.

\medskip\noindent
\emph{AI disclosure.} %
{\bfseries All definitions, statements, and proofs in this paper have been obtained and written by the human authors.}

Claude Opus 5.5 and GPT-6 Astra were used for technical editing of the manuscript:
both were used for polishing the language, copying parts of the text and adapting them to new contexts,
proofreading,
and completing bibliographic data. 
All parts of the paper including the bibliography were verified by the authors,
who retain full responsibility for the text.

\phantomsection
\addcontentsline{toc}{section}{References}
\bibliographystyle{alpha} %
\newcommand{\etalchar}[1]{$^{#1}$}

\appendix

\section{Uniform extension to arbitrary tropical axioms}
\label{sec:appendix-generalized}

The resolution axioms play no special role in the simulations of Section~\ref{sec:main-results}: the same arguments identify the static, treelike dynamic and daglike systems over any \emph{unary} class of tropical axioms. This appendix defines the generalized systems and extends results to them.

\subsection{Tropical axioms and the generalized systems}

\begin{definition}[Tropical axiom]
\label{def:tropical-axiom}
A \emph{tropical axiom} is a tropical inequality over the dual-encoded variables,
\begin{equation}
\label{eq:general-axiom}
\min_{j \in [\kappa]} \big(a_j^T x - b_j\big) \;\leq\; c^T x - d,
\qquad a_j, c \in \Z_+^{2N},\ b_j, d \in \Z,
\end{equation}
that is valid at every Boolean point (the points with $x_i \in \{0, 1\}$ and $\overline{x_i} = 1 - x_i$); by the reduction to single right-hand terms (Section~\ref{sec:tropical-arithmetic}) we assume the form~\eqref{eq:general-axiom} with one term on the right. The standard axioms, the resolution axioms and the division axioms of Definition~\ref{def:division-axioms} are tropical axioms; the clause encodings~\eqref{eq:encoding-i} are tropical inequalities but, for nontrivial clauses, not axioms. The size of a tropical axiom is measured with its constants $b_j, d$ and the entries of the vectors $a_j, c$ encoded in unary, as in Section~\ref{sec:notation}; a \emph{class} $\mathcal{C}$ of tropical axioms is \emph{unary} if the sizes of its axioms, so measured, are bounded by a polynomial in $N$. A class may be of superpolynomial size; we only require that for a concrete CNF one can exhibit a polynomial-size subset of $\mathcal{C}$, and a refutation may use axioms only from this subset.
\end{definition}

\begin{definition}[Generalized static and treelike systems]
\label{def:generalized-mp}
Let $\mathcal{C}$ be a class of tropical axioms. \emph{Generalized {\mpnsr} over $\mathcal{C}$} (resp.\ \emph{generalized treelike {\mppcr} over $\mathcal{C}$}) is the system {\mpns} (resp.\ treelike {\mppc}) whose premises are the clause encodings~\eqref{eq:encoding-i} of the refuted CNF, the standard axioms of Definition~\ref{def:mpnsr}, and the axioms of $\mathcal{C}$. The systems {\mpnsr} and treelike {\mppcr} are the special case $\mathcal{C} = \{\min(x_i, \overline{x_i}) \leq 0 : i \in [N]\}$, the class of resolution axioms.
\end{definition}

\begin{definition}[Daglike system over $\mathcal{C}$]
\label{def:daglike-tropical}
The \emph{daglike system over $\mathcal{C}$} operates on threshold conditions $f \geq s$ with arbitrary integer thresholds. A proof is a directed acyclic graph with the single source $0 \geq 0$ in which every non-source line is obtained by the {\rulecl} rule of Definition~\ref{def:thrres} or by the \emph{application in a context} (in the sense of Lemma~\ref{lm:rules-are-applications}) of a standard axiom or of an axiom~\eqref{eq:general-axiom} of $\mathcal{C}$:
\[
\frac{f + a_1^T x \geq s + b_1 \quad \cdots \quad f + a_{\kappa}^T x \geq s + b_{\kappa}}{f + c^T x \geq s + d}\,.
\]
A refutation is a proof containing the line $0 \geq 1$; size and unarity are as in Definition~\ref{def:thrres}. By Lemma~\ref{lm:rules-are-applications}, {\thrres} is exactly the daglike system over the class of resolution axioms. Soundness is proved as in Theorem~\ref{thm:thrres-sound}: adding the common context $f - s$ to both sides of an axiom preserves its validity at every Boolean point.
\end{definition}

\subsection{The generalized statements}

Here generalized {\mpnsr} and treelike {\mppcr} use binary encoding, whereas the generalized daglike threshold system is unary. The restriction to unary axiom classes bounds the sizes of the axioms, not the encoding of tropical certificates or proofs.

\begin{theorem}[Generalized Main Lemma]
\label{thm:generalized-main}
Let $\mathcal{C}$ be a unary class of tropical axioms. Generalized {\mpnsr} over $\mathcal{C}$ and the daglike system over $\mathcal{C}$ polynomially simulate each other. Quantitatively: a generalized {\mpnsr} refutation of size $S$ yields a daglike refutation of size $\poly(S, W, N)$, where $W \geq 1$ bounds the absolute values of the integer constants $b_j, d$ of the axioms used in the refutation, and a daglike refutation of size $S$ yields a generalized {\mpnsr} refutation of size $\poly(S)$.
\end{theorem}

\begin{theorem}[Generalized daglike to treelike]
\label{thm:generalized-daglike-to-treelike}
Let $\mathcal{C}$ be a unary class of tropical axioms. Generalized treelike {\mppcr} over $\mathcal{C}$ polynomially simulates the daglike system over $\mathcal{C}$.
\end{theorem}

\begin{theorem}[Generalized flattening]
\label{thm:generalized-treelike-to-ns}
For every class $\mathcal{C}$ of tropical axioms, generalized {\mpnsr} over $\mathcal{C}$ polynomially simulates generalized treelike {\mppcr} over $\mathcal{C}$.
\end{theorem}
\begin{proof}
Theorem~4.1 of~\cite{AGH25} is stated for an arbitrary set of premises; apply it to the premise set of Definition~\ref{def:generalized-mp}.
\end{proof}

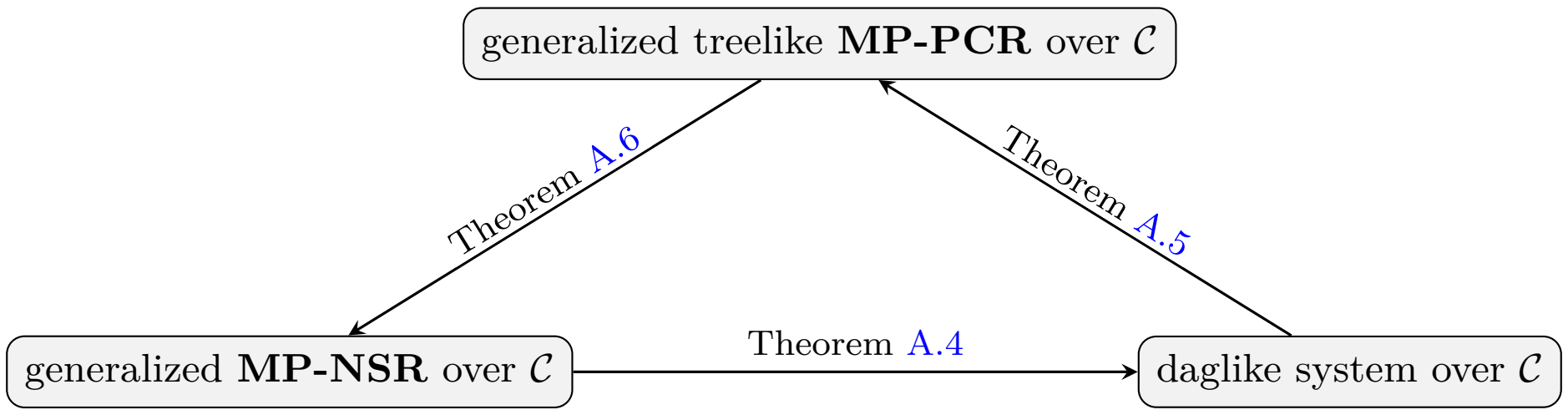
\begin{figure}[ht]
\centering
\begin{tikzpicture}[x=1cm, y=1cm]
\node[sys] (gt) at (0, 2.6)  {generalized treelike {\mppcr} over $\mathcal{C}$};
\node[sys] (gn) at (-4.2, 0) {generalized {\mpnsr} over $\mathcal{C}$};
\node[sys] (gd) at (4.2, 0)  {daglike system over $\mathcal{C}$};
\draw[sim] (gt) -- (gn) node[lab, midway, above, sloped] {Theorem~\ref{thm:generalized-treelike-to-ns}};
\draw[sim] (gn) -- (gd) node[lab, midway, above=2pt] {Theorem~\ref{thm:generalized-main}};
\draw[sim] (gd) -- (gt) node[lab, midway, above, sloped] {Theorem~\ref{thm:generalized-daglike-to-treelike}};
\end{tikzpicture}
\caption{The equivalence triangle over an arbitrary unary class $\mathcal{C}$ of tropical axioms (Theorem~\ref{thm:triangle}); an arrow from $\Pi_1$ to $\Pi_2$ means that $\Pi_2$ polynomially simulates $\Pi_1$. Enlarging the class moves the entire equivalence class up in strength.}
\label{fig:triangle-cycle}
\end{figure}

\begin{theorem}[Equivalence triangle]
\label{thm:triangle}
For every unary class $\mathcal{C}$ of tropical axioms, generalized {\mpnsr} over $\mathcal{C}$, generalized treelike {\mppcr} over $\mathcal{C}$ and the daglike system over $\mathcal{C}$ polynomially simulate one another, by the cycle of Figure~\ref{fig:triangle-cycle}.
\end{theorem}

\begin{corollary}[Unary coefficients, generalized]
\label{cor:generalized-unary}
If generalized {\mpnsr} over $\mathcal{C}$ has a refutation of size $S$ using axioms with integer constants bounded by $W$ in absolute value, then it has a refutation of size $\poly(S, W, N)$ in which all coefficients are rationals with a common denominator $\poly(S, N)$ and unary size $\poly(S, W, N)$.
\end{corollary}

\subsection{What changes in the proofs}
\label{sec:what-changes}

\paragraph{Daglike to treelike.} The proof of Theorem~\ref{thm:daglike-to-treelike} manipulates a rule application only through its description as the application of a premise $\min_{r}(a_r^T x - b_r) \leq c^T x - d$ in a context, and Definition~\ref{def:daglike-tropical} builds the generalized system directly from such applications. The invariant $V_i \colon 0 \leq \bigoplus_{j\leq i}T_j$ and the substitution step of that proof apply to these applications unchanged, which proves Theorem~\ref{thm:generalized-daglike-to-treelike}; the only quantitative change is that the coefficients of the treelike derivation are now bounded in terms of the constants of the axioms of $\mathcal{C}$, which for a unary class keeps the derivation unary.

\paragraph{The certificate--graph characterization.} Theorem~\ref{thm:refut-subgraph-equiv} and the positive-cycle potential lemma (Lemma~\ref{lm:potentials}) are stated for an arbitrary finite set of tropical inequalities and need no change. Corollary~\ref{cor:generalized-unary}, the generalized form of Corollary~\ref{cor:unary-coefficients}, follows by the same argument, with the edge weights of the generalized encoding bounded by $W$ in absolute value.

\paragraph{The normal form.} A \emph{generalized refutation strategy graph} for $F$ and $\mathcal{C}$ is a strategy graph (Definition~\ref{def:strategy-graph}) in which every vertex is either a {\rulecl} vertex of Table~\ref{tab:local-types}, or an \emph{axiom vertex}: for a standard axiom or an axiom~\eqref{eq:general-axiom} of $\mathcal{C}$ and a position $v \geq c$ coordinate-wise, it carries the outgoing edges
\[
v \;\longrightarrow\; v - c + a_j \quad \text{of weight } d - b_j, \qquad j \in [\kappa]
\]
(for the standard and resolution axioms these are the last four rows of Table~\ref{tab:local-types}). The proof of the normal-form lemma (Lemma~\ref{lm:normal-form}) computes the local structure of the monomial graph for an arbitrary premise with a single right-hand term, so it applies with this type inventory as is; only the bounds are revised: the edge weights are now at most $W' \leq 2W$ in absolute value and, for a unary class, the position change along an edge is at most $A = \poly(N)$ in $\ell_1$-norm.

\paragraph{The backward reading.} An axiom vertex at the position $v \geq c$, labeled $f_v \leq t$, has the children labeled $f_{v - c + a_j} \leq t - d + b_j$; rewriting $f_v = f_{v-c} + c^T x$, this is precisely the backward application of the axiom~\eqref{eq:general-axiom} in the context $f_{v-c} - (t - d)$, i.e., the reversal of the application rule of Definition~\ref{def:daglike-tropical}. The reversal proposition (Proposition~\ref{prop:reversal}) transfers with the same one-line comparison per type, and all resulting positions are nonnegative exponent vectors because the applications require $v \geq c$.

\paragraph{The remaining constructions.} The positive-cycle layering lemma (Lemma~\ref{lm:layering}) is purely graph-theoretic and is reused without modification, with $W' \leq 2W$. The cleanup lemma (Lemma~\ref{lm:cleanup}) uses only the standard axioms, which are present in every generalized system. The unfolding and compression arguments (Lemma~\ref{lm:unfolding-lemma} and Proposition~\ref{prop:compression}) inspect a vertex only through its reading as a backward move, so they apply with the generalized type inventory; the out-degrees are now bounded by the maximal branching $\kappa$ of the axioms, and the thresholds of the unfolded proof are $O(rW)$, where $r$ is the number of vertices of the strategy graph. The assembly of Theorem~\ref{thm:generalized-main} then repeats Section~\ref{sec:assembly} with the parameters $W' \leq 2W$ and $A = \poly(N)$ bounding both the clause widths and the vectors $a_j, c$ of the (unary) axioms, giving the size $\poly(S, W, N)$ in the forward direction and $\poly(S)$ in the converse one.

For unary axiom classes, $W$ is polynomially bounded in $N$; we retain it explicitly to display the dependence on axiom constants. Note that the size of the unfolded proof is polynomial in the numerical value of $W$, not in its bitsize: the construction is pseudopolynomial for axioms with binary constants, which is why the generalized statements are restricted to unary classes.

\end{document}